\documentclass[letterpaper,usenatbib]{mn2e}
\usepackage{amsmath}
\usepackage[breaklinks, colorlinks, citecolor=blue]{hyperref}
\usepackage{epsfig}
\usepackage{multirow}
\usepackage{times}
\usepackage{verbatim}
\def\gtsim{~\rlap{$>$}{\lower 1.0ex\hbox{$\sim$}}}
\def\ltsim{~\rlap{$<$}{\lower 1.0ex\hbox{$\sim$}}}

\title[PAH excitation in nearby spiral galaxies]{Polycyclic aromatic hydrocarbon excitation in nearby spiral galaxies}

\author[G. J. Bendo et al.]
    {G. J. Bendo$^1$, N. Lu$^{2,3}$, A. Zijlstra$^4$ \newauthor \\
    $^1$   UK ALMA Regional Centre Node, Jodrell Bank Centre for Astrophysics,
           Department of Physics and Astronomy, University of Manchester, 
           Oxford Road,\\ Manchester M13 9PL, United Kingdom\\
    $^2$   National Astronomical Observatories, Chinese Academy of Sciences
           (CAS), Beijing 100012, People's Republic of China\\
    $^3$   China-Chile Joint Center for Astronomy, Camino El Observatorio 1515,
           Las Condes, Santiago, Chile\\
    $^4$   Jodrell Bank Centre for Astrophysics,
           Department of Physics and Astronomy, University of Manchester, 
           Oxford Road,\\ Manchester M13 9PL, United Kingdom\\
}
\date{}
\pagerange{\pageref{firstpage}--\pageref{lastpage}}
\pubyear{}

\begin{document}
\label{firstpage}
\maketitle

\begin{abstract}
We have examined polycyclic aromatic hydrocarbon (PAH) excitation in a sample of 25 nearby face-on spiral galaxies using the ratio of mid-infrared PAH emission to dust mass.  Within 11 of the galaxies, we found that the PAH excitation was straightforwardly linked to ultraviolet or mid-infrared star formation tracers, which, along with other results studying the relation of PAH emission to star formation, indicates that the PAHs are most strongly excited in dusty shells around the star forming regions.  Within another 5 galaxies, the PAH emission is enhanced around star forming regions only at specific galactocentric radii.  In 6 more galaxies, PAH excitation is more strongly correlated with the evolved stellar populations as traced by 3.6~$\mu$m emission.  The results for the remaining 3 galaxies were ambiguous.   The radial gradients of the PAH/dust ratios were generally not linked to log(O/H) gradients except when the log(O/H) gradients were relatively steep.  Galaxies in which PAHs were excited by evolved stars had relatively high far-ultraviolet to mid-infrared ratios, indicating that variations in the link between PAH excitation and different stellar populations is linked to changes in dust attenuation within galaxies.  Alternately, differences in morphology could make it more likely that PAHs are excited by evolved stars, as 5 of the 6 galaxies where this occurs are late-type flocculent spiral galaxies.  These heterogeneous results demonstrate the complexity of describing PAH excitation and have broad implications for using PAH emission as a star formation tracer as well as for modelling dust emission and radiative transfer.
\end{abstract}


\section{Introduction}
\label{s_intro}
\addtocounter{footnote}{4}

The interstellar medium (ISM) emits a series of broad spectral features in the mid-infrared that are commonly attributed to polycyclic aromatic hydrocarbons \citep[PAHs; see][for a review]{tielens08}\footnote{We acknowledge that other particles could produce these spectral features \citep[for example, see ][]{jones13}, but we will refer to the producers of these emission lines as PAHs.}.  PAHs are effectively transitory between interstellar molecules and larger carbonaceous dust grains, so they can be used as a tracer of interstellar dust and gas.  Additionally, PAHs are often assumed to be excited primarily by ultraviolet photons and are sometimes expected to be associated with star formation activity \citep{tielens08}, although it is possible that PAHs can be excited by lower energy photons as well \citep{li02}.  

Early studies using data from the Infrared Space Observatory \citep[ISO; ][]{kessler96} found that PAH emission in other galaxies was associated with other star formation tracers, particularly optical and near-infrared hydrogen line emission and mid-infrared thermal dust emission \citep{roussel01, forsterschreiber04}, although some evidence was found for PAHs being excited by older stars in diffuse regions \citep{boselli04}.  However, these analysis were limited by the telescope's angular resolution and sensitivity.  Later studies using data from the {\it Spitzer} Space Telescope \citep{werner04} examined the relation between PAH emission and other star formation tracers more closely and found that the PAH emission functioned very poorly as a star formation tracer on scales of a kpc or smaller.  In star forming regions, PAH emission appeared suppressed relative to other star formation tracers, while in diffuse regions, PAH emission appeared high relative to the other tracers \citep{calzetti05, prescott07, bendo08}.  Shell-like features in PAH emission were identified around both galactic and extragalactic star forming regions \citep{helou04, bendo06, povich07, kirsanova08, watson08, dewangan13}, illustrating that PAH emission is suppressed within the centres of photodissociation regions.  Additionally, PAH emission was found to decline relative to  24~$\mu$m  hot dust continuum emission in low metallicity regions \citep{engelbracht05, engelbracht08, madden06, calzetti07, gordon08}.  This apparent relation with metallicity has been attributed to increases in the hardness and intensity of the radiation field that is a consequence of the low-metallicity regions containing less dust, but alternative explanations have been put forward, such as variations in dust grain formation in low metallicity environments \citep{galliano08} or PAH destruction by shocks related to enhanced supernova activity in low metallicity systems \citep{ohalloran06}.  Having said this, some fraction of PAH emission can still be associated with other star formation tracers like H$\alpha$ emission \citep{crocker13}.  PAH emission is still used as an indicator of star formation activity, particularly in metrics used to distinguish between starburst and AGN activity \citep{genzel98, peeters04, dale06}, and globally-integrated PAH emission has been calibrated as a star formation tracer for nearby spiral galaxies and similar systems \citep{zhu08, kennicutt09, shipley16}.

However, the detection of large-scale diffuse PAH emission throughout the Milky Way by the Cosmic Background Explorer \citep[e.g. ][]{dwek97} and by {\it Spitzer} \citep[e.g. ][]{lu04} suggested that part of the PAH emission within galaxies originates from regions outside star forming regions.  Indeed, a stronger association has been found between PAH emission and dust emission at $\geq$100~$\mu$m.  Early analyses using millimetre and submillimetre data found a stronger correlation between PAH emission and 850~$\mu$m emission than between PAH emission and hot dust emission \citep{mattila99, haas02}.  Later analyses with data from {\it Spitzer} demonstrated that PAH emission was strongly correlated with 160~$\mu$m emission \citep{bendo08, verley09} and that the PAH/160~$\mu$m surface brightness ratio depended upon the 160~$\mu$m surface brightness \citep{bendo08}.  Since dust model results based on {\it Spitzer} data indicated that most of the 160~$\mu$m emission should originate from dust heated by the diffuse interstellar radiation field (ISRF) from the total stellar populations in these galaxies \citep{draine07b}, the expectation was that the PAHs were also excited by the diffuse ISRF and that the strength of the PAH emission relative to cold dust emission was tied to the intensity of the ISRF.

Initial results from the {\it Herschel} Space Observatory \citep{pilbratt10} presented a more confusing picture of the nature of the relation between PAH and far-infrared dust emission.  Analyses of M33 by \citet{calapa14} and of NGC~2403 by \citet{jones15} demonstrated that the PAH/250~$\mu$m ratio in these two galaxies correlates well with 3.6~$\mu$m surface brightnesses, indicating that the PAHs in these galaxies are excited by the diffuse ISRF.  However, \citet{jones15} also presented an analysis of M83 that demonstrated that the PAH/250~$\mu$m ratio in that galaxy was enhanced along the spiral arms but that the ratio peaked downstream from the photoionizing regions themselves.  This indicated that the PAHs in M83 are primarily excited by a younger stellar population but that the relation is more complicated than just localized PAH excitation.  Additionally, \citet{lu14} presented an analysis of M81 that identified 67\% of the PAH excitation being attributed to evolved stars and the remainder being attributed to a young stellar population.  

\begin{table*}
\centering
\begin{minipage}{165mm}
\caption{Position, dimension, morphology, and nuclear activity information for sample galaxies}
\label{t_sample}
\begin{tabular}{@{}lcccccccc@{}}
\hline
Galaxy &
  Right &
  Declination$^a$ &
  \multicolumn{2}{c}{Galaxy Disc$^b$} &
  Inclination$^c$ &
  Morpho- &
  Nucleus &
  Nucleus Type \\
&
  Ascension$^a$ &
  (J2000) &
  Dimensions &
  Position &
  ($^{\circ}$) &
  logical  &
  Type$^e$ &
  Reference \\
&
  (J2000) &
  &
  (arcmin) &
  Angle ($^{\circ}$) &
  &
  Type$^d$ &
  &
  \\
\hline
NGC 300 &
  00:54:53.5 &
  -37:41:04 &
  21.5 $\times$ 15.2 &
  115 &
  46 &
  SAd &
  SF &
  \citet{payne04}$^f$ \\
NGC 628 &
  01:36:41.7 &
  +15:47:01 &
  11.5 $\times$ 9.7 &
  118$^h$ &
  7$^g$ &
  SAc &
  SF &
  \citet{moustakas10} \\
NGC 925 &
  02:27:16.9 &
  +33:34:45 &
  10.5 $\times$ 5.9$^d$ & 
  102$^d$ & 
  66$^g$ & 
  SABd &
  SF &
  \citet{moustakas10} \\
NGC 1097 &
  02:46:19.0 &
  -30:16:30 &
  12.1 $\times$ 7.6 &
  128 &
  54 &
  SBb &
  AGN &
  \citet{moustakas10} \\
NGC 1365 &
  03:33:36.4 &
  -36:08:25 &
  12.8 $\times$ 7.8 &
  42 &
  55 & 
  SBb &
  SF/AGN &
  \citet{alonsoherrero12} \\
NGC 2403 &
  07:36:51.4 &
  +65:36:09 &
  21.9 $\times$ 12.3$^d$ & 
  127$^d$ & 
  63$^g$ &
  SABcd &
  SF &
  \citet{moustakas10} \\
NGC 3031 &
  09:55:33.2 &
  +69:03:55 &
  27.0 $\times$ 15.6 &
  155 &
  59$^g$ &
  SAab &
  AGN &
  \citet{moustakas10} \\
NGC 3184 &
  10:18:16.8 &
  +41:25:27 &
  8.2 $\times$ 7.0 &
  138$^h$ &
  16$^g$ & 
  SABcd &
  SF &
  \citet{moustakas10} \\
NGC 3351 &
  10:43:57.7 &
  +11:42:14 &
  8.4 $\times$ 6.2 &
  11 &
  41$^g$ &
  SBb &
  SF &
  \citet{moustakas10} \\
NGC 3621 &
  11:18:16.5 &
  -32:48:51 &
  12.3 $\times$ 7.1$^d$ & 
  159$^d$ &
  65$^g$ &
  SAd &
  AGN &
  \citet{moustakas10} \\
NGC 3938 &
  11:52:49.4 &
  +44:07:15 &
  5.6 $\times$ 5.1 &
  21 &
  25 &
  SAc &
  AGN &
  \citet{moustakas10} \\
NGC 4051 &
  12:03:09.6 &
  +44:31:53 &
  6.7 $\times$ 5.0 &
  125 &
  43 & 
  SABbc &
  AGN &
  \citet{ho97} \\
NGC 4254 &
  12:18:49.6 &
  +14:24:59 &
  6.4 $\times$ 5.2 &
  84 &
  37 & 
  SAc &
  SF/AGN &
  \citet{moustakas10} \\
NGC 4303 &
  12:21:54.9 &
  +04:28:25 &
  7.6 $\times$ 6.3 &
  143 &
  35 &  
  SABcd &
  SF &
  \citet{ho97} \\
NGC 4321 &
  12:22:54.8 &
  +15:49:19 &
  10.4 $\times$ 8.0 &
  174 &
  41 & 
  SABbc &
  AGN &
  \citet{moustakas10} \\
NGC 4501 &
  12:31:59.1 &
  +14:25:13 &
  9.2 $\times$ 5.0 &
  142 &
  59 &  
  SAc &
  AGN &
  \citet{ho97} \\
NGC 4548 &
  12:35:26.4 &
  +14:29:47 &
  7.3 $\times$ 6.2 &
  145 &
  33 & 
  SBc &
  AGN &
  \citet{ho97} \\
NGC 4579 &
  12:37:43.5 &
  +11:49:05 &
  8.4 $\times$ 6.0 &
  96 &
  46 &  
  SABb &
  AGN &
  \citet{moustakas10} \\
NGC 4725 &
  12:50:26.6 &
  +25:30:03 &
  11.5 $\times$ 8.1 &
  30 &
  48 & 
  SABab &
  AGN &
  \citet{moustakas10} \\
NGC 4736 &
  12:50:53.0 &
  +41:07:14 &
  19.0 $\times$ 15.0 &
  116 &
  41$^g$ &
  SAab &
  AGN &
  \citet{moustakas10} \\
NGC 5055 &
  13:15:49.3 &
  +42:01:45 &
  17.3 $\times$ 10.5 &
  107 &
  59$^g$ &
  SAbc &
  AGN &
  \citet{moustakas10} \\
NGC 5236 &
  13:37:00.9 &
  -29:51:56 &
  19.0 $\times$ 17.7 &
  47 &
  24$^g$ &
  SABc &
  SF &
  $^i$ \\
NGC 5248 &
  13:37:32.0 &
  +08:53:07 &
  7.0 $\times$ 5.3 &
  104 &
  42 &  
  SABbc &
  SF &
  \citet{ho97} \\
NGC 5457 &
  14:03:12.5 &
  +54:20:56 &
  23.6 $\times$ 17.4 &
  56 &
  18$^g$ &
  SABcd &
  SF &
  \citet{ho97} \\
NGC 7793 &
  23:57:49.8 &
  -32:35:28 &
  10.9 $\times$ 6.9 &
  100 &
  50$^g$ &
  SAd &
  SF &
  \citet{moustakas10} \\
\hline
\end{tabular}
$^a$ These data are taken from the NASA/IPAC Extragalactic Database (http://ned.ipac.caltech.edu). \\
$^b$ These data are taken from \citet{sheth10} unless otherwise specified.  Position angles are from north through east.\\
$^c$  Unless otherwise specified, inclinations are calculated using $\theta_{inc} = \sin^{-1} ( [1-(r_{maj}/r_{min})^{-2} ]^{0.5} / [1-10^{(-0.86-0.106T)}]^{0.5} )$, which is taken from the HyperLeda website (http://leda.univ-lyon1.fr/) and which is a variant of an equation derived by \citet{hubble26}.  In this equation, $r_{maj}$ and $r_{min}$ are the radii of the major and minor axes, and $T$ is the numerical value corresponding to the morphological type specified by RC3. \\
$^d$ These data are taken from RC3. \\
$^e$ Nuclei are divided into star forming nuclei (SF) and active galactic nuclei (AGN) following the convention given by \citet{moustakas10}.\\
$^f$ The identification of the nucleus of NGC~300 as star forming is based on the absence any AGN-like central X-ray or radio sources as found by \citet{payne04}.\\
$^g$ These inclination are based on H{\small I} measurements compiled by \citet{walter08}, which should be more accurate than the standard inclination calculation.\\ 
$^h$ These position angles apply to the isophote for the 3.6~$\mu$m emission as given by \citet{sheth10}.  However, \citet{walter08} lists position angles (applicable to the orientation of the gas discs) of 20$^\circ$ for NGC~628 and 179$^\circ$ for NGC~3184, which are significantly different.  For defining the optical disc in the analysis, the position angles in this table will be used, but for calculating galactocentric radii, the \citet{walter08} values will be used.\\
$^i$ Information from multiple references were used to label the nucleus as SF.  See the text in Section~\ref{s_sample} for more information.
\end{minipage}
\end{table*}

These four galaxies combined present an inconsistent picture of exactly how PAHs are excited and how PAHs are related to cold dust.  To confuse the interpretation further, \citet{bendo15} found not only that the dust emitting at 160~$\mu$m and 250~$\mu$m emission could be heated by either a young stellar population or evolved stars but also that the relative proportion of dust heated by evolved stars was difficult to predict for any galaxy.  These results on dust heating contradicted the model results from \citet{draine07b} used in the interpretation of the how the ratio of PAH/160~$\mu$m correlated with 160~$\mu$m surface brightness as presented by \citet{bendo08}.

Hence, the exact picture of how PAHs are excited is unclear.  If PAH emission is going to be used as a tracer of either star formation or cold dust, then the nature of how PAHs are excited needs to be understood better.  Additionally, if dust emission and radiative transfer models are going to try to accurately reproduce the infrared spectral energy distributions (SEDs) of galaxies, then correctly linking PAHs to their excitation source is important.

In this paper, we use the ratio of PAH emission to dust masses calculated from {\it Herschel} data to explore how PAH emission is enhanced relative to cold dust emission in a broad sample of nearby face-on spiral galaxies.  PAH emission by itself is a function of both the PAH excitation and PAH surface density, so any correlations between PAH emission and far-infrared dust emission could arise either because the PAHs are excited by the same stellar population that heats the dust or because the PAH surface density is enhanced in the same locations where dust surface density is enhanced.  We can use the PAH/dust ratio as a metric of PAH excitation in which the dependence on surface density has been removed.  This assumes that PAHs are mainly associated with cold interstellar dust, as has been shown in several studies \citep[e.g. see ][]{bendo08, verley09, jones15, cortzen19}.  In comparing the PAH/dust ratio to 3.6~$\mu$m emission, which is used as a tracer of the evolved stellar populations, and to 154~nm and 24~$\mu$m emission, which are used as tracers of the young stellar populations, we can identify how PAHs are generally excited.

\section{Sample}
\label{s_sample}

We used multiple criteria to select the sample for this analysis.  We started with all Sa-Sd galaxies listed in the Third Reference Catalogue of Bright Galaxies \citep[RC3; ][]{devaucouleurs91} that are outside the Local Group, that have major axes larger than 5~arcmin, and that have major/minor axis ratios less than 2.  A total of 83 galaxies match these criteria.  We excluded the two strongly interacting galaxy pairs where it would have been difficult to disentangle stellar or interstellar emission from the two galaxies (M51 and NGC 3226/3227).  After this, we selected the galaxies observed by {\it Spitzer} at 3.6, 4.5, 5.8, 8.0, and 24~$\mu$m and observed by {\it Herschel} at 250 and 350~$\mu$m.  This reduced the sample to 36 galaxies.  We then removed galaxies where the 8~$\mu$m image is severely saturated in the centre (NGC~1068, 1808, and 2146), mainly because the combination of the saturated pixels themselves and the associated image artefacts would cause multiple problems when analyzing the data.  We also excluded galaxies where the data detected at the $3\sigma$ level after the binning process described in Section~\ref{s_data_convreb} has an area smaller than the area of a 3$\times$3~arcmin square (NGC~1512, 1566, 3338, 4450, 4826, and 4939).  In these galaxies with relatively compact PAH and dust emission, it is difficult to see spatial variations in the PAH/dust ratio at the 25~arcsec resolution of the 350~$\mu$m data or to relate the spatial variations in the PAH/dust ratio to stellar populations or galactic substructures.  Finally, we did not use galaxies within 15$^\circ$ of the plane of the Milky Way Galaxy (IC~342 and NGC~6946) where multiple foreground stars cause problems with using the near- and mid-infrared data and where the high dust extinction cause problems with using ultraviolet data as a star formation tracer.

The resulting sample contains 25 galaxies.  The sample list as well as position, dimension, morphology, and nuclear activity information galaxy information is provided in Table~\ref{t_sample}.  Distances as well as the effective scales of the 24~arcsec subregions used in our analysis (see Section~\ref{s_data_convreb}) are listed in Table~\ref{t_dist}.

A significant fraction of the sample was observed with {\it Spitzer} as part of {\it Spitzer} Infrared Nearby Galaxies Survey \citep[SINGS; ][]{kennicutt03}  or the Local Volume Legacy (LVL) survey \citep{dale09}, but other galaxies were not covered in those survey.  Similarly, most of the galaxies were observed  with {\it Herschel} as part of the {it Herschel}-SPIRE Local Galaxies Guaranteed Time Programs (which included the Very Nearby Galaxies Survey and the Herschel Reference Survey \citep[HRS; ][]{boselli10}), the {\it Herschel} Virgo Cluster Survey \citep[HeViCS; ][]{davies10}, or Key Insights on Nearby Galaxies: A Far-Infrared Survey with Herschel \citep[KINGFISH; ][]{kennicutt11}.  However, some of our sample galaxies were not covered in any of these surveys.

AGN could potentially affect PAH emission, which is why we listed nuclear activity information in Table~\ref{t_sample}.  PAH emission has generally been observed to be weaker relative to infrared continuum emission in AGN, although it is unclear whether this is caused by the destruction of PAHs \citep{siebenmorgen04, wu09, monfredini19} or the dilution of the line emission by the continuum, as PAHs are detected in the circumnuclear environments around AGN and could be excited by those AGN (e.g. \citealt{alonsoherrero14, esquej14, jensen17}; but also see \citealt{esparzaarredondo18}).  The nuclear activity designations are preferentially taken from the survey by \citet{moustakas10}, with \citet{ho97} used as the reference for most other galaxies.  NGC~300, NGC~1365, and NGC~5236 are not listed in either of these surveys, so we relied on other references for nuclear activity designations.  NGC~300 has no apparent strong nuclear X-ray or radio source \citep[e.g. ][]{payne04}, so we treat its nucleus as dominated by star formation.  NGC~1365 has a well-documented composite starburst/AGN for its nucleus \citep[e.g ][]{sharp10, alonsoherrero12}.  At least one AGN candidate has been detected in the centre of NGC~5236 \citep{long14}, but given the uncertainty of this identification \citep{yukita16}, the relatively low X-ray brightness of the candidate compared to other X-ray sources in the galaxy, and the number of young stellar clusters found in the centre of the galaxy \citep[e.g. ][]{harris01}, we treat the nucleus as dominated by star formation.

When determining which distances to list in Table~\ref{t_dist}, preference is generally given to most recent distances determined using Cepheids, the tip of the red giant branch method, or supernovae.  \citet{tully13} list distances based on a combination of these methods, which made the measurements preferable to others based on just one method.  When no such measurements are available, preference is given to the most recent calculation based on the Tully-Fisher relation.

\section{Data processing and preparation}

\begin{table}
\caption{Distance information for sample galaxies}
\label{t_dist}
\begin{center}
\begin{tabular}{@{}lccc@{}}
\hline
Galaxy &
  Distance &
  Scale of &
  Reference \\
&
  (Mpc)  &
  24 arcsec bin &
  \\
&
  &
  (kpc) &
  \\
\hline
NGC 300 &
  1.98 $\pm$ 0.05 &
  0.23 &
  \citet{tully13} \\
NGC 628 &
  9.0 $\pm$ 0.6 &
  1.05 & 
  \citet{dhungana16} \\
NGC 925 &
  8.9 $\pm$ 0.3 &
  1.04 &
  \citet{tully13} \\
NGC 1097 &
  24.8 $\pm$ 0.6 &
  2.89 &
  \citet{rodriguez14} \\
NGC 1365 &
  17.8 $\pm$ 0.8 &
  2.07 &
  \citet{tully13} \\
NGC 2403 &
  3.18 $\pm$ 0.08 &
  0.37 &
  \citet{tully13} \\
NGC 3031 &
  3.61 $\pm$ 0.10 &
  0.42 &
  \citet{tully13} \\
NGC 3184 &
  14.4 $\pm$ 0.3 &
  1.68 &
  \citet{ferrarese00} \\
NGC 3351 &
  10.5 $\pm$ 0.3 &
  1.22 &
  \citet{tully13} \\  
NGC 3621 &
  6.73 $\pm$ 0.19 &
  0.78 &
  \citet{tully13} \\
NGC 3938 &
  22.1 $\pm$ 1.7 &
  2.57 &
  \citet{rodriguez14} \\
NGC 4051 &
  8.8 $\pm$ 1.7 &
  1.02 &
  \citet{sorce14} \\
NGC 4254 &
  14.6 $\pm$ 1.9 &
  1.70 &
  \citet{poznanski09} \\
NGC 4303 &
  20.0 $\pm$ 0.6 &
  2.33 &
  \citet{rodriguez14} \\
NGC 4321 &
  13.9 $\pm$ 0.4 &
  1.62 &
  \citet{tully13} \\
NGC 4501 &
  15.4 $\pm$ 0.4 &
  1.79 &
  \citet{mandel11} \\
NGC 4548 &
  17.1 $\pm$ 0.6 &
  1.99 &
  \citet{tully13} \\
NGC 4579 &
  21.0 $\pm$ 1.9 &
  2.44 &
  \citet{ruizlapuente96} \\
NGC 4725 &
  12.6 $\pm$ 0.5 &
  1.47 &
  \citet{tully13} \\
NGC 4736 &
  4.59 $\pm$ 0.17 &
  0.53 &
  \citet{tully13} \\
NGC 5055 &
  9.0 $\pm$ 0.4 &
  1.05 &
  \citet{tully13} \\
NGC 5236 &
  4.66 $\pm$ 0.15 &
  0.54 &
  \citet{tully13} \\
NGC 5248 &
  14 $\pm$ 3 &
  1.63 &
  \citet{tully16} \\
NGC 5457 &
  6.95 $\pm$ 0.19 &
  0.81 &
  \citet{tully13} \\
NGC 7793 &
  3.58 $\pm$ 0.12 &
  0.42 &
  \cite{tully13} \\
\hline
\end{tabular}
\end{center}
\end{table}

\label{s_data}

\begin{table*}
\centering
\begin{minipage}{160mm}
\caption{Data properties overview}
\label{t_dataoverview}
\begin{tabular}{@{}lccccc@{}}
\hline
Wavelength &
  Telescope &
  Image Pixel &
  PSW FWHM &
  Calibration &
  References \\
($\mu$m) &
  /Instrument &
  Size (arcsec) &
  (arcsec) &
  Uncertainty &
  \\
\hline
0.154 &
  GALEX &
  1.5 &
  4.2 &
  4.5\% &
  \citet{morrissey07} \\
3.6 &
  {\it Spitzer}/IRAC &
  0.75 &
  1.7 &
  10\%$^a$ &
  \citet{irac15} \\
4.5 &
  {\it Spitzer}/IRAC &
  0.75 &
  1.7 &
  10\%$^a$ &
  \citet{irac15} \\
5.8 &
  {\it Spitzer}/IRAC &
  0.75 &
  1.9 &
  10\%$^a$ &
  \citet{irac15} \\
8.0 &
  {\it Spitzer}/IRAC &
  0.75 &
  2.0 &
  10\%$^a$ &
  \citet{irac15} \\
24 &
  {\it Spitzer}/MIPS &
  1.5 &
  6 &
  4\% &
  \citet{engelbracht07} \\
70 &
  {\it Herschel}/PACS &
  1.6 &
  6$^b$ &
  5\% &
  \citet{balog14}, \citet{lutz15}\\
100 &
  {\it Herschel}/PACS &
  1.6 &
  7$^b$ &
  5\% &
  \citet{balog14}, \citet{lutz15}$^c$ \\
160 &
  {\it Herschel}/PACS &
  3.2 &
  12$^b$ &
  5\% &
  \citet{balog14}, \citet{lutz15}$^c$ \\
250 &
  {\it Herschel}/SPIRE &
  6 &
  18 &
  4\% &
  \citet{bendo13}, \citet{valtchanov18}$^d$ \\
350 &
  {\it Herschel}/SPIRE &
  8 &
  24 &
  4\% &
  \citet{bendo13}, \citet{valtchanov18}$^d$ \\
\hline
\end{tabular}
$^a$ This uncertainty is primarily from the surface brightness correction.  The photometric accuracy is otherwise 3\%.\\
$^b$ The PSF sizes quoted here are for the 20 arcsec s$^{-1}$ scan speeds.  At higher scan speeds, the PSFs appear more elongated.\\
$^c$ Available from https://www.cosmos.esa.int/documents/12133/996891/PACS+photometer+point+spread+function .\\
$^d$ Available from https://www.cosmos.esa.int/documents/12133/1035800/The+Herschel+Explanatory+Supplement\%2C\%20Volume+IV+-+THE+SPECTRAL+AND+PHOTOMETRIC+IMAGING+RECEIVER+\%28SPIRE\%29 . \\
\end{minipage}
\end{table*}

{\it Spitzer} 8.0~$\mu$m images are used to trace the PAH emission in the sample galaxies.  However, stars and interstellar hot dust produce a small fraction of the total emission in the band, so we also use 3.6, 4.5, 5.8, and 24~$\mu$m data to remove these additional sources of emission.

For dust surface densities, we use values calculated from measurements in multiple {\it Herschel} bands.  Measurements from a single waveband at $>$160~~$\mu$m could be used to trace dust surface density.  The data would sample primarily large grain emission from the Rayleigh-Jeans side of the SED, and the use of such empirical data would not require performing any additional calculations.  However, measurements in the individual {\it Herschel} bands are still significantly dependent on temperature variations.  For dust temperatures ranging from 10 to 35~K, which could be expected for subregions within nearby spiral galaxies \citep[e.g. ][]{bendo15}, the surface brightness per unit dust mass could vary by $\gtsim$75$\times$ at 250~$\mu$m, $\gtsim$25$\times$ at 350~$\mu$m, and $\gtsim$10$\times$ at 500~$\mu$m.  

We therefore decided to use measurements from multiple {\it Herschel} bands to calculate dust surface densities.  While multiple radiative transfer or dust emission models could be fit to the data, not all models are publicly available, and these dust models are sufficiently complex to apply that modelling an individual nearby galaxy could be the subject of a paper by itself.  Additionally, the models often have their own inherent assumptions about the dust emission, and either these assumptions or the model results may not accurately describe the separation between diffuse dust heated by the interstellar radiation field and dust heated locally by star forming regions \citep{bendo15}.  Single modified blackbodies are far easier to apply to data and could be appropriate to use when the dust seen in the far-infrared is primarily heated by a single stellar population \citep[e.g. ][]{bendo10} but not when the dust seen at different wavelengths is heated by different sources \citep[e.g. ][]{bendo15}.  This is particularly problematic when the modified blackbody is scaled by a power law $\lambda^\beta$ where the index $\beta$ is a free parameter in the fit \citep{kirkpatrick14, hunt15}.  However, we have found that a single modified blackbody fitted to just the {\it Herschel} 250 and 350~$\mu$m bands can provide masses for the large dust grains that are within 10\% of the actual masses for many plausible dust emission scenarios and within 20\% for all but some extreme situations.  A demonstration of this is illustrated in Appendix~\ref{a_dustmass}.  In practice, though, the modified blackbody fit to the 250 and 350~$\mu$m bands sometimes supersedes what is measured at shorter wavelengths, so we also use {\it Herschel} 70, 100, and 160~$\mu$m to check this and further constrain the dust SED when appropriate.  More details are given in Section~\ref{s_data_addcalc}.

We used the 3.6~$\mu$m band as a tracer of the evolved stellar population, as the emission in this band originates primarily from the Rayleigh-Jeans side of the stellar SEDs and as the band is relatively unaffected by dust extinction.  Some careful analyses of the mid-infrared colours of nearby galaxies have indicated that the band could also include hot dust emission \citep[e.g. ][]{mentuch09, mentuch10, meidt12}.  However, \citet{bendo15} found that using 1.6~$\mu$m data instead of 3.6~$\mu$m data made very little difference when studying evolved stellar populations as a dust heating source in their study, implying that dust emission in the 3.6~$\mu$m band was negligible for most regions in their sample galaxies.  The same would probably be true for our analysis, especially since we use many of the same galaxies that appear in the \citet{bendo15} sample. 

Ultraviolet continuum, H$\alpha$ line, and mid-infrared continuum data are the star formation tracers that are broadly available for galaxies in general.  Physically, each of these tracers are related to star formation in slightly different ways.  The H$\alpha$ emission traces unobscured photoionizing stars with ages $\ltsim$5 Myr.  The ultraviolet emission traces unobscured young photoionizing and non-photoionizing blue stars and therefore may trace populations with ages up to or over 100 Myr, although the mean age of the stars detected in these bands is $\sim$10~Myr \citep{kennicutt12}.  The mid-infrared emission comes from hot dust that has empirically been associated with H$\alpha$ emission and other tracers of photoionizing stars within individual galaxies \citep[e.g. ][]{calzetti05, calzetti07, prescott07, bendo15}.  If PAH excitation is connected to star forming regions, then PAHs may be expected to be connected with mid-infrared emission than the other star formation tracers given that the PAHs are treated as part of a continuum between dust grains and interstellar molecules.  On the other hand, given the results from \citet{jones15} showing that PAHs were excited most strongly in regions in M83 offset from the peaks in the H$\alpha$ and 24~$\mu$m emission but where the ultraviolet emission was still strong, it would make sense to also compare PAH excitation to ultraviolet emission.  

From a practical standpoint, working with ultraviolet and mid-infrared data is much easier than H$\alpha$ data.  The Galaxy Evolution Explorer \citep[GALEX; ][]{martin05} and {\it Spitzer} have produced data with relatively uniform properties that cover all of the galaxies in this sample.  In contrast, H$\alpha$ data tends to be heterogeneous, as multiple telescopes and instruments may be used to acquire the data and as the observing conditions may vary among the observations, even among data compiled by individual surveys.   Furthermore, the creation of H$\alpha$ images typically involves subtracting the stellar continuum emission, which can leave artefacts in the data such as incompletely-subtracted or oversubtracted stellar continuum emission from foreground stars and galaxy bulges as well as irregularities in the image backgrounds, and the presence of the [N{\small II}] lines also complicates measurements from H$\alpha$ images.  Additionally, many of the publicly-available H$\alpha$ images either do not completely cover the optical discs of the target galaxies or do not include substantial background regions for measuring background noise.  For some galaxies, no H$\alpha$ data are publicly available.

Given this, we use GALEX 154~nm and {\it Spitzer} 24~$\mu$m data as star formation tracers for all galaxies in our sample.  We refer to publicly-released H$\alpha$ images where appropriate, but we do not incorporate the data into our quantitative analyses.  The 154~nm and 24~$\mu$m data have their own limitations, as both can include a small fraction of emission from evolved stars and as the 24~$\mu$m band may include emission from dust heated by evolved stars, but these problems are relatively minor.

Table~\ref{t_dataoverview} provides a summary of the characteristics of the images in each waveband used in this analysis.  This table includes information and references regarding the FWHMs of the PSFs and the calibration uncertainties of each instrument.

\subsection{GALEX 154~nm data}

\begin{table}
\caption{Sources of GALEX and {\it Spitzer}-MIPS data}
\label{t_data_galexmips}
\begin{center}
\begin{tabular}{@{}lcc@{}}
\hline
Galaxy &
  GALEX &
  {\it Spitzer} 24~$\mu$m \\
&
  Tile Name &
  Data Source \\
\hline
NGC 300 &
  GI1\_061002\_NGC0300 &
  LVL \\
NGC 628 &
  GI3\_050001\_NGC628 &
  SINGS \\
NGC 925 &
  NGA\_NGC0925 &
  SINGS \\
NGC 1097 &
  NGA\_NGC1097 &
  SINGS \\
NGC 1365 &
  FORNAX\_MOS06 &
  [this paper] \\
NGC 2403 &
  GI3\_050002\_NGC2403 &
  \citet{bendo12b} \\
NGC 3031 &
  GI1\_071001\_M81 &
  \citet{bendo12b} \\
NGC 3184 &
  AIS\_86 &
  SINGS \\
NGC 3351 &
  NGA\_NGC3351 &
  SINGS \\
NGC 3621 &
  NGA\_NGC3621 &
  SINGS \\
NGC 3938 &
  AIS\_102 &
  SINGS \\
NGC 4051 &
  AIS\_102 &
  [this paper] \\
NGC 4254 &
  GI2\_017001\_J121754p144525 &
  \citet{bendo12b} \\
NGC 4303 &
  NGA\_NGC4303 &
  \citet{bendo12b} \\
NGC 4321 &
  GI5\_057004\_NGC4312 &
  \citet{bendo12b} \\
NGC 4501 &
  AIS\_223 &
  \citet{bendo12b} \\
NGC 4548 &
  GI2\_034006\_Malin1 &
  \citet{bendo12b} \\
NGC 4579 &
  NGA\_Virgo\_MOS07 &
  \citet{bendo12b} \\
NGC 4725 &
  AIS\_219 &
  \citet{bendo12b} \\
NGC 4736 &
  NGA\_NGC4736 &
  SINGS \\
NGC 5055 &
  NGA\_NGC5055 &
  SINGS \\
NGC 5236 &
  GI3\_050007\_NGC5236 &
  \citet{bendo12b} \\
NGC 5248 &
  AIS\_221 &
  \citet{bendo12b} \\
NGC 5457 &
  GI3\_050008\_NGC5457 &
  LVL \\
NGC 7793 &
  NGA\_NGC7793 &
  SINGS \\
\hline
\end{tabular}
\end{center}
\end{table}

For ultraviolet continuum data, we used pipeline processed images taken with the FUV (154~nm) filter from GALEX Release 6 and 7\footnote{Accessible at http://galex.stsci.edu/GR6/ .}, which are preferable to the data taken with the GALEX NUV (227~nm) filter because they will tend to sample emission from younger stars better \citep{kennicutt12}.  For each galaxy, we selected the image with the longest exposure time in the FUV band as long as the centre of the target galaxy was two times the radius of the optical disc from the edge of the frame.  The positioning constraint ensured not only that the entire galaxy falls within the field of view but also that the data contain enough background data on all sides of the galaxy for measuring the background.  Most of the sources were observed in targeted observations as part of either specific GALEX surveys, such as the GALEX Nearby Galaxy Survey \citep[which was later incorporated into the GALEX Ultraviolet Atlas of Nearby Galaxies; see ][]{gildepaz07}, or as guest investigator programs.  However, some sources are only covered by the relatively shallow All-sky Imaging Survey \citep{bianchi17}.  Although the exposure times are typically $\sim$10$\times$ shorter for the AIS compared to most other observations, the signal-to-noise of the data after the application of the convolution and rebinning steps described in Section~\ref{s_data_convreb} is sufficient for our analysis.  The specific data frames used in the analysis are listed in Table ~\ref{t_data_galexmips}.

\subsection{{\it Spitzer} 3.6-8.0 $\mu$m data}

3.6, 4.5, 5.8, and 8.0~$\mu$m images from the Infrared Array Camera \citep[IRAC; ][]{fazio04} have been publicly released by SINGS, the LVL survey, and Spitzer Survey of Stellar Structure in Galaxies \citep{sheth10} for most of our sample galaxies.  However, these images were created using older basic calibrated data frames (BCDs).  The {\it Spitzer} archive now contains additional corrected basic calibrated data frames (CBCDs) that include corrections for stray light, saturation, column pulldown, banding, and muxbleed-related effects.  We found that we could use the CBCDs to create better images than the ones publicly available, so we chose to create new 3.6, 4.5, 5.8, and 8.0~$\mu$m images of all of the sample galaxies based on the newer CBCDs.  We used only data obtained during the cryogenic mission (when the telescope had coolant) since these data were sufficient for imaging all of every galaxy's disc and since these data are less noisy than data available from the warm mission (after the telescope depleted its coolant).

As the first step in creating the images, we identified any data frames where the IRAC pipeline did not completely remove flatfield effects and applied a flatfield correction based on data covering background regions.  Next, for all galaxies, we identified and subtracted the median background from pixels in the individual data frames that did not lie within the galaxies' optical discs or cover other bright sources.  For frames lying entirely within the optical discs of some of the larger galaxies, we determined the background levels by interpolating among adjacent frames.  After background subtraction, we mosaicked the data using slightly modified version of the default imaging scripts within the MOsaicker and Point source EXtractor \citep[{\sc MOPEX}; ][]{makovoz05} version 18.5.  The image output settings in the {\sc Mosaic Settings} and {\sc Fiducial Image Frame} modules are set to produce images with pixel scales of 0.75~arcsec and with the images oriented so that north is up and east is to the left.  In the {\sc Med Filter} module used to remove large scale structure for the outlier rejection modules, we used {\sc SExtractor} background filtering with a filter size of 50 pixels, which should allow for better removal of local surface brightness variations.   The interpolation method in the {\sc Mosaic Interpolate} module is set to "Drizzle" with a drizzle factor of 1 because this is an effective way to resample pixelated data from the individual data frames into a final mosaic with a different orientation and different pixel size.  Finally, in the {\sc Detect} module, the detection threshold is set to 3$\sigma$ to be more sensitive to outliers than the default of 4$\sigma$.  

After imaging the data, we applied the surface brightness correction factors listed in the IRAC Instrument Handbook\footnote{Available from http://irsa.ipac.caltech.edu/data/SPITZER/docs/irac/ iracinstrumenthandbook/IRAC\_Instrument\_Handbook.pdf .} \citep{irac15} and subtracted any residual background emission from the final images.  Although the photometric calibration for point sources is expected to be accurate to within 3\%, we quote a calibration uncertainty of 10\% in Table~\ref{t_dataoverview} based on the uncertainties in the surface brightness corrections.

\subsection{{\it Spitzer} 24~$\mu$m data}

24~$\mu$m images for most galaxies in the sample have already been created from Multiband Imaging Photometer for {\it Spitzer} \citep[MIPS; ][]{rieke04} data by SINGS, the LVL survey, and \citet{bendo12b} for the SAG2 and HeViCS surveys.  All of these data were created using the same pixel scale and the same flux calibration specified by \citet{engelbracht07}.  Moreover, all of these images were created from raw data processed with the MIPS Data Analysis Tools \citep[MIPS DAT; ][]{gordon05}.  Even though slightly different processing steps were used by \citet{bendo12b} compared to the final data releases from the other surveys, the quality of the resulting images are effectively the same.  We therefore used these data in our analysis.  Table~\ref{t_data_galexmips} lists the specific source that we used for each 24~$\mu$m image.

24~$\mu$m images based on {\it Spitzer} data for NGC 1365 and NGC~4051 have not been publicly released to our knowledge, so we created new mosaics of these data using all available data from the {\it Spitzer} archive.  The image of NGC 1365 was created from raw archival data frames using MIPS DAT version 3.10 \citep{gordon05} as well as the additional tools described by \citet{bendo12b}.  However, because of technical problems with the MIPS DAT, we used enhanced BCDs from the archive and {\sc MOPEX} version 18.5 to create the 24~$\mu$m image of NGC~4051.  Before mosaicking, we applied the latent image removal, mirror-position-independent flatfield, and zodiacal light removal steps described by \citet{bendo12b}.  When mosaicking, we used the default {\sc MOPEX} imaging script but with changes to create an image with 1.5 arcsec pixels and with north up and east to the left.

\subsection{{\it Herschel} 70-350~$\mu$m data}

The {\it Herschel} Photodetector Array Camera and Spectrometer \citep[PACS; ][]{poglitsch10} acquired images of all of the sample galaxies at 160~$\mu$m and at either 70~$\mu$m, 100~$\mu$m, or both 70 and 100~$\mu$m depending on the target.  The Spectral and Photometric Imaging Receiver \citep[SPIRE; ][]{griffin10} imaged all of the sample galaxies at 250, 350, and 500~$\mu$m.  Some of these images have been publicly released by SAG2, HeViCS, and KINGFISH for many but not all of the galaxies in this sample.  However, many of the SPIRE images and some of the PACS images were created before the final calibration tables were released.  Additionally, for the publicly-released SPIRE images, some of the mapmaking tools that were used introduced subtle changes in the data compared to standard SPIRE mapmakers \citep[see ][ for a discussion]{xu13}\footnote{Available from http://herschel.esac.esa.int/twiki/pub/Public/ \\ SpireDocsEditableTable/report\_main\_v5.pdf .}.

In the case of the PACS data, we chose to download the most recent PACS images available in the {\it Herschel} archive (Standard Product Generation version 14.2.0), as we were unable to create better images ourselves.  We selected the Level 3 data products, which are images created from all observations of the same field, when these were available.  We otherwise used the Level 2.5 data products, which are created from cross-scanned pairs of observations.  In cases where sources were observed with both the scan map and parallel (PACS and SPIRE) map modes, we selected the scan map data because the observations were generally deeper.  When possible, we selected images that were created using the {\sc Unimap} mapmaking algorithm \citep{piazzo15}, as these data had the lowest noise levels.  Otherwise, we used images created using the {\sc Scanamorphos} mapmaker \citep{roussel13}, which produces maps that are comparable to the ones from {\sc Unimap} \citep{paladini13, paladini14}\footnote{Available from https://www.cosmos.esa.int/documents/12133/996891/ \\ PACS+Map-making+Tools+-+Analysis+and+Benchmarking and https:// \\ www.cosmos.esa.int/documents/12133/996891/PACS+Map-making+Tools+-+Update+on+Analysis+and+Benchmarking .}.

In the case of the SPIRE data, we were able to recreate images that were better for our purposes than what was available in the archive.  In particular, we preferred to use images with finer 350~$\mu$m pixel scales than what is provided by the archive, and we also had our own preferences for calibrating the data for extended emission.  We used all available archival SPIRE data for each target, including small scan maps, large scan maps, and parallel (PACS and SPIRE) maps.  All data were reprocessed using the {\it Herschel} Interactive Processing Environment \citep{ott10} version 15 and SPIRE calibration tree spire\_cal\_14\_3.  The bolometer timeline data were first processed through standard pipeline scripts that include electrical crosstalk correction, cosmic ray removal, low pass filter response correction, flux conversion, temperature drift correction, and bolometer time response correction steps.  After this, extended gain corrections were applied to the timeline data to correct the detectors' response to extended sources, which is optimal for performing photometry on nearby galaxies.  For each field with parallel mode maps, which typically cover areas of skies with widths of several degrees, we trimmed the timelines to data only covering areas within 0.5 degrees of the source.  The standard baseline removal and destriping programs were applied to remove signal offsets among the individual detector timelines, and then the timeline data were converted into maps with the standard naive mapmaker using a pixel scale of 6~arcsec at 250~$\mu$m and 8~arcsec at 350~$\mu$m.

\subsection{Image convolution and rebinning}
\label{s_data_convreb}

The analysis techniques used in our work are similar to what was used in related prior analyses of the relation of PAH emission to cold dust \citep{bendo08, jones15} as well as related analyses on dust heating \citep{bendo10, bendo12a, bendo15}.  These techniques are based on comparing measurements of the PAH/dust ratios to star formation tracers and 3.6~$\mu$m emission within subregions of individual galaxies that are equivalent in size to the FWHM of the PSF and therefore should be statistically independent.  To do this comparison, though, we must match the angular resolutions of all of the images to the data with the largest PSF, which are the 350~$\mu$m data, or else our analysis will contain artefacts related to the differing beam sizes. For quantitative comparisons, we also need to bin the data so that we are working with data that are close to statistically independent.

We first identified locations with foreground stars (typically identified as unresolved sources in the 3.6-8.0~$\mu$m data with $I_{3.6\mu m}/I_{8.0\mu m}>5$) and removed the stars by interpolating over the pixels.  Next, we matched the PSFs of all data to the PSF of the {\it Herschel} 350~$\mu$m data using the convolution kernels published by \citet{aniano11}\footnote{Available for download from https://www.astro.princeton.edu/ \\ $\sim$ganiano/Kernels.html .}.  When applied to the images, these convolution kernels not only match the FWHMs of the PSFs to the PSF of the 350~$\mu$m images but also match axisymmetric substructures of the PSFs, such as Airy rings, to the substructures in the 350~$\mu$m PSF.  After this, the images were regridded into the same coordinate system with the same pixel scale as the 350~$\mu$m images.  These data are sufficient for creating maps of the PAH/dust ratios and for performing qualitative comparisons of these ratio maps to images of either 154~nm, 3.6~$\mu$m, or 24~$\mu$m emission.  Each pixel will be 8~arcsec, so  approximately 3 pixels span the 25~arcsec FWHM of the data, which should be sufficient for display purposes.  For quantitative analyses, we rebinned the data into 24~arcsec bins.  This bin size was selected because it is an integer number of map pixels that is similar in size to the FWHM of the 350~$\mu$m data.  The measurements for the individual bins should be statistically independent except for very bright point sources.  Using a smaller pixel size would result in subsampling the PSFs of individual sources, and since the subsampled data points should all have the same SEDs, the data could appear artificially correlated.

In both the binned and unbinned data, we masked pixels where at least 25\% of the pixel was affected by foreground stars in one or more bands.  We also manually masked binned data that are strongly affected by residual muxbleed effects in the 8.0~$\mu$m data.  For making images of the ratio of PAH emission to dust mass, we used data within the  galaxies' optical discs (as defined in Table~\ref{t_sample}) where the PAH emission and dust mass were determined to the $\geq$3$\sigma$ level.  For selecting binned data for quantitative analyses, we used data measured at the $\geq$3$\sigma$ level in PAH emission and dust mass as well as in the 154~nm, 3.6~$\mu$m, and 24~$\mu$m bands.

The 24~arcsec bins in our analysis cover spatial scales that vary by an order of magnitude between nearby and more distant galaxies.  We do not attempt to compare the slopes of relations obtained from different galaxies, as the variation in distance could affect the slopes of the data.  Our analysis relies upon seeing differences among the relations measured within individual galaxies, and as long as differences can be seen between the structures in the star formation tracers, the 3.6~$\mu$m emission, and the PAH/dust ratios, we should be able to indicate which stellar populations are more strongly connected to PAH excitation.  However, in more distant galaxies, the star forming regions and evolved stellar populations will appear more blended.  \citet{bendo15}, which used similar binned data for nearby galaxies in their analysis, tested how their results changed when simulating galaxies at different distances, and they found that their results were robust against distance effects.  We replicated that analysis with our data for NGC~3031 and NGC~5457, the two galaxies in our sample with the largest angular sizes, and found that our conclusions would not change.  A detailed discussion is given in Appendix~\ref{a_disttest}.  Given this, we do not expect our results to be biased by spatial resolution effects.

\begin{figure*}
\epsfig{file=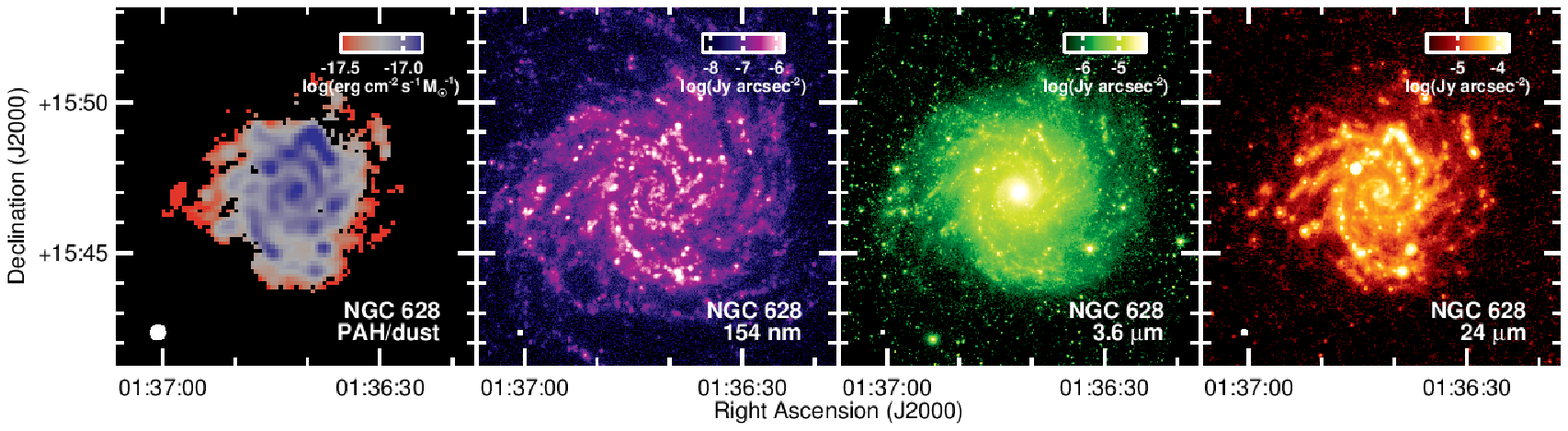}

\vspace{0.3cm}

\epsfig{file=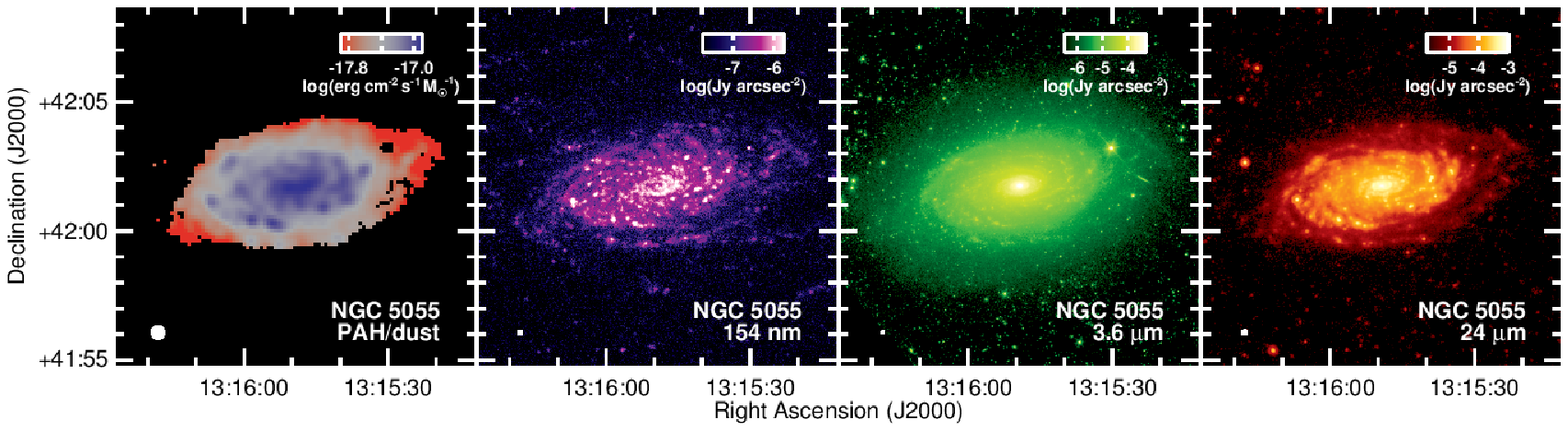}

\caption{Maps of the PAH/dust ratios, 154~nm emission, 3.6~$\mu$m emission, and 24~$\mu$m emission in two galaxies where the PAH excitation (as traced by the PAH/dust ratios) is connected to star formation activity (as traced by either the 154~nm emission, 24~$\mu$m emission or the combination of these).  All plots are scaled logarithmically, with the units given in the colour bar in the upper right corner of each image.  The images of the dust surface densities used to create the PAH/dust ratio maps have been smoothed by a 3 $\times$ 3 pixel (24 $\times$ 24~arcsec) top-hat function.  Pixels not detected at the 3$\sigma$ level in PAH emission or dust mass (before smoothing) are coloured black.  The PSF for each image is shown as a white circle in the lower left corner of each image.  The PAH/dust ratio maps are shown at a resolution of 24~arcsec, while the other maps have PSFs corresponding to the FWHM in Table~\ref{t_dataoverview}.}
\label{f_map_sfr}
\end{figure*}

\begin{figure*}
\epsfig{file=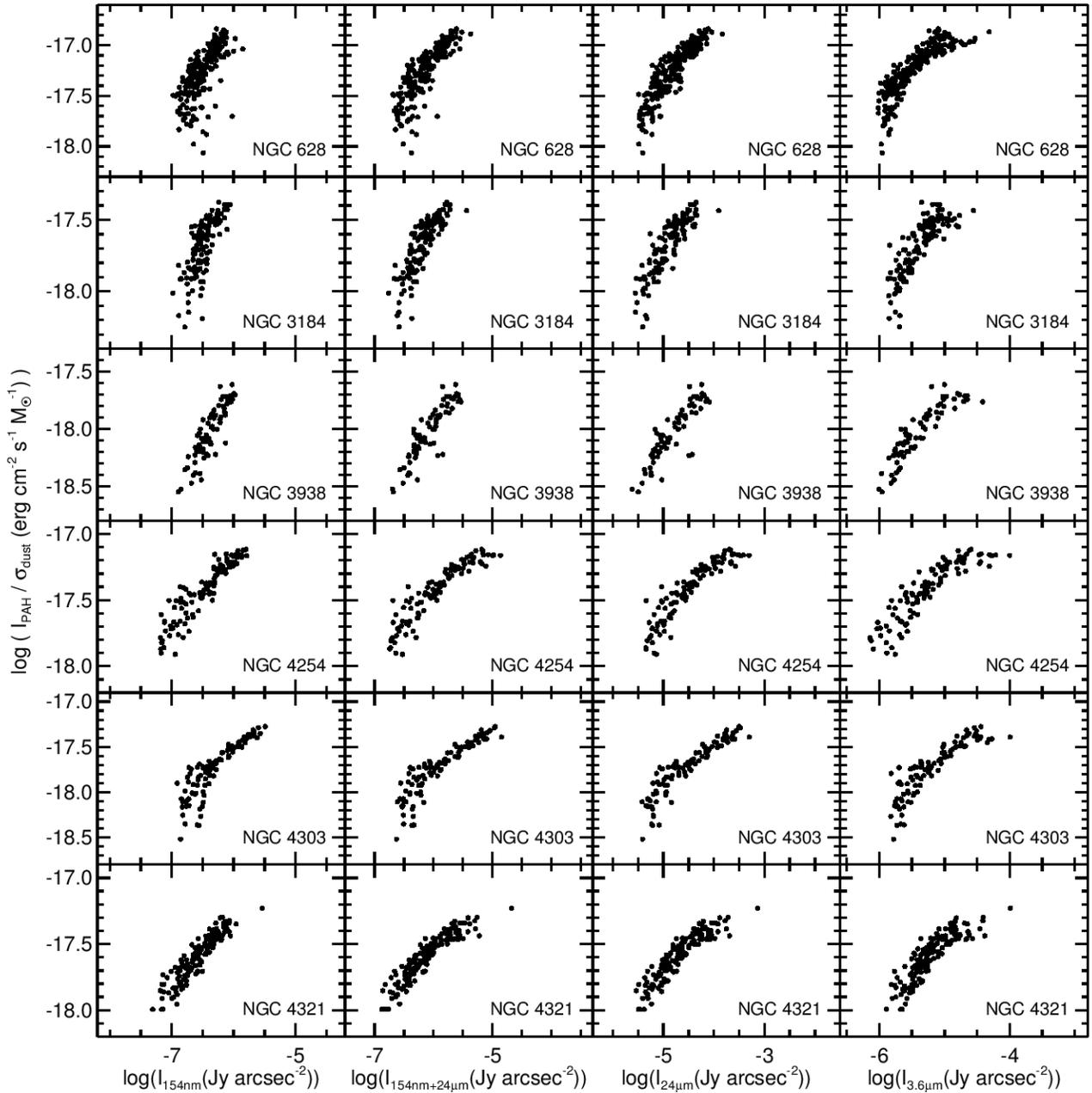,width=17.5cm}
\caption{Plots of the PAH/dust ratios versus star formation tracers and versus 3.6~$\mu$m emission as measured within 24~arcsec square bins for the galaxies where the ratios are most strongly correlated with one or more of the star formation tracers.  The uncertainties in the log($I_{PAH}/\sigma_{dust}$) values vary between high- and low-surface-brightness regions, but the average uncertainties are $\sim$0.05. Uncertainties in the x-axis values are typically smaller than the data points.  Statistics for the relations are in Table~\ref{t_comprat_sfr}.}
\label{f_comprat_sfr}
\end{figure*}

\addtocounter{figure}{-1}
\begin{figure*}
\epsfig{file=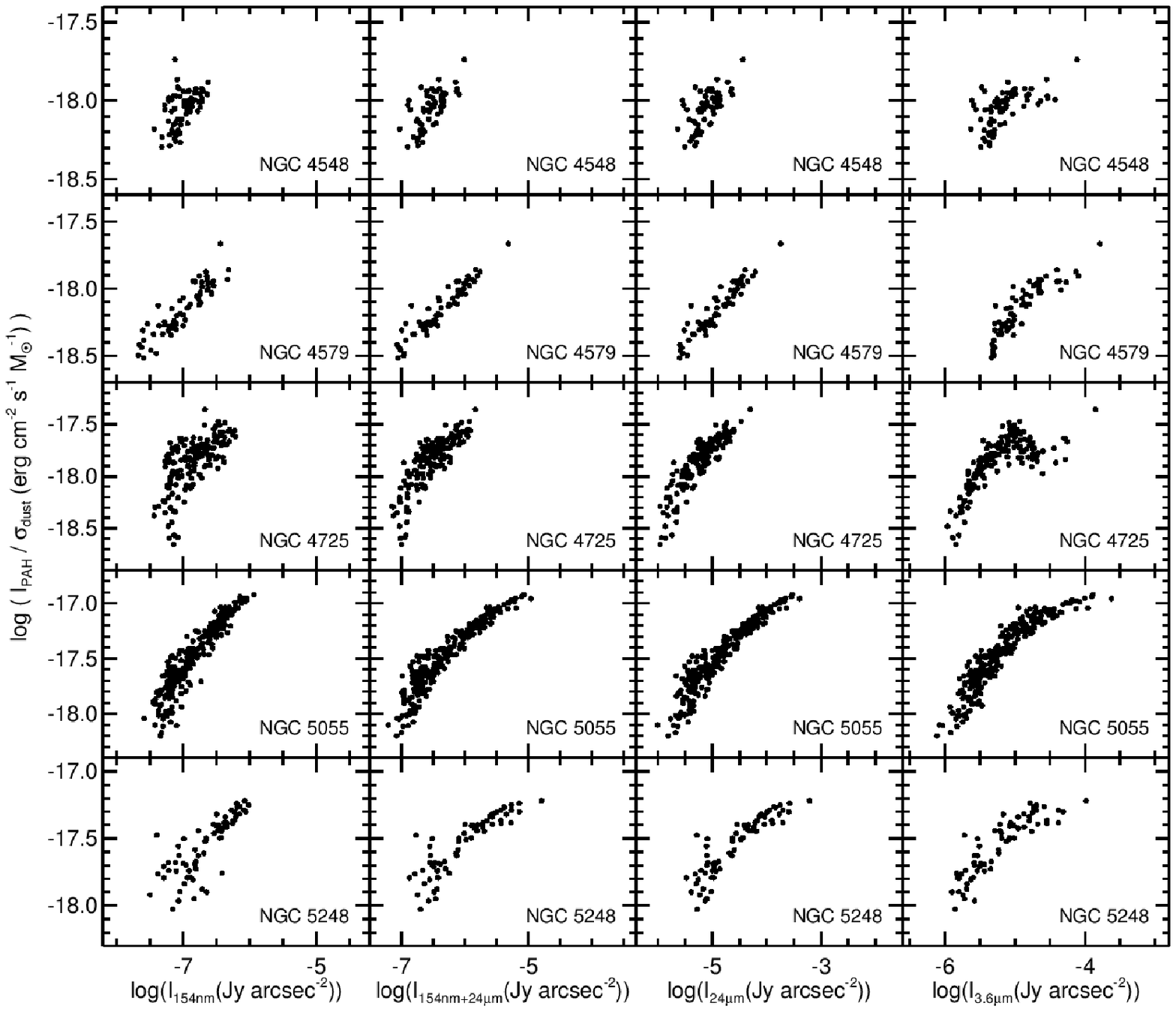,width=17.5cm}
\caption{Continued.}

\end{figure*}

\subsection{Additional calculations}
\label{s_data_addcalc}

While the 3.6 and 4.5~$\mu$m bands are expected to dominated by stellar emission and the 5.8, 8.0, and 24$\mu$m bands are expected to be dominated by interstellar PAH or hot dust emission, stellar emission may be detectable up to 24~$\mu$m.  Multiple techniques have been developed for subtracting the stellar emission from the longer wavelength bands.  We used
\begin{equation}
I_{4.5\mu m ~ SCS} = I_{4.5\mu m} - 0.60 I_{3.6\mu m}
\end{equation}
\begin{equation}
I_{5.8\mu m ~ SCS} = I_{5.8\mu m} - 0.40 I_{3.6\mu m}
\end{equation}
\begin{equation}
I_{8.0\mu m SCS} = I_{8.0\mu m} - 0.246 I_{3.6\mu m}
\end{equation}
\begin{equation}
I_{24\mu m ~ SCS} = I_{24\mu m} - 0.033 I_{3.6\mu m}
\end{equation}
from \citet{jones15}.  In these equations, “SCS” stands for stellar continuum subtracted.  The application of these equations relies on the assumption that the 3.6~$\mu$m originates from stellar emission and that dust extinction is negligible in these bands.  The typical changes are $\sim$10\% for the 8.0~$\mu$m data and $\sim$2\% for the 24~$\mu$m data.  It is known that the 3.6~$\mu$m band may also contain interstellar dust and PAH emission \citep{lu03, lu04, mentuch09, mentuch10, meidt12}.  The emission may contribute $\sim$10\% of the global 3.6~$\mu$m emission, although the contribution could be higher in locations with very strong star formation activity.  This will add some uncertainties to our correction, but it should be at levels of $\ltsim$1\% for the 8.0 and 24~$\mu$m bands.

To calculate the PAH emission at 8~$\mu$m, which consists of a complex of multiple spectral emission features, we used
\begin{equation}
\begin{split}
I_{PAH} = 1.43\times10^{-10} \ \ \ \ \ \ \ \ \ \ \ \ \ \ \ \ \ \ \ \ \ \ \ \ \ \ \ \ \ \ \ \ \ \ \ \ \ \ \ \ \ \ \ \ \ \ \ \ \ \ \ \ \ \ \ \ \ \ \ \ \\
  \times [I_{8.0\mu m ~ SCS} - (0.091 + 0.314 I_{8.0\mu m}/I_{24\mu m}) \\
  \times (I_{4.5\mu m ~ SCS}+I_{5.8\mu m ~ SCS})^{0.718}I_{24\mu m SCS}^{0.282}]
\end{split}
\label{e_pahcalc}
\end{equation}
derived by \citet{marble10} based on an analysis of the SEDs of galaxies in the LVL Survey.  This equation is expected to be accurate to within 6\%.  However, it may produce unreliable results in cases where an AGN is the predominant infrared emission source (as is the case in NGC~3031, NGC~4051, and NGC~4725). Also note that, even though 24~$\mu$m data is included in this calculation, the relation between $I_{PAH}$ and $I_{24\mu m SCS}$ is very similar to the relation between $I_{8.0\mu m ~ SCS}$ and $I_{24\mu m SCS}$, which demonstrates that the use of Equation~\ref{e_pahcalc} does not introduce an artificial correlation between PAH and 24~$\mu$m emission.

All far-ultraviolet data presented in this paper is corrected for foreground dust extinction from the Milky Way Galaxy using
\begin{equation}
I_{154nm~corr} = I_{154nm~obs} e^{7.9E(B-V)/1.086},
\end{equation}
which was derived by \citet{lee11}.  The $E(B-V)$ values are computed by the NASA/IPAC Extragalactic Database\footnote{Accessible at http://ned.ipac.caltech.edu/ .} using the analysis results from \citet{schlafly11}.  

While we perform separate comparisons between the PAH/dust ratio and either of the star formation tracers, we also found it useful to use a combined far-ultraviolet and mid-infrared star formation tracer given by 
\begin{equation}
I_{154nm+24\mu m} = I_{154nm}+3.89 \left( \frac{0.154~\mu\mbox{m}}{24~\mu\mbox{m}} \right)I_{24\mu m}.
\end{equation}
This is effectively the equation derived by \citet{hao11} for using 24~$\mu$m emission to correct the 154~nm GALEX band for dust extinction intrinsic to the target galaxies.  However, we will refer to it as the 154~nm~+~24~$\mu$m metric to distinguish it from the version of the 154~nm not corrected for intrinsic dust extinction (but still corrected for foreground dust extinction).  Some of the 24~$\mu$m emission may be unassociated with star formation, so the application of this correction is imperfect.  The locations with the most egregious problems are the centres of a few galaxies where the predominant infrared source is an AGN or where significant mid-infrared emission is produced by dust unassociated with star formation seen in ultraviolet or H$\alpha$ data \citep[such as in NGC~3031, NGC~4548, and NGC~4725; see][]{bendo15}.  The extinction correction based on the 24~$\mu$m data should be reasonably accurate for most other subregions within our sample galaxies.

Dust surface densities $\sigma_{dust}$ for each pixel and 24~arcsec bin are determined by fitting
\begin{equation}
\sigma_{dust}=\frac{I_\nu D^2}{\kappa_\nu B_\nu (T)}
\label{e_dustmass}
\end{equation}
to the data.  In this equation, which is based on other equations originally derived by \citet{hildebrand83}, $D$ is the distance, $\kappa_\nu$ is the emissivity function, and $B_\nu (T)$ is the blackbody function.  For $\kappa_\nu$, we interpolate among the values given by \citet{draine03}, which closely approximate but differ slightly from an emissivity function proportional to $\nu^2$.  To calculate $\sigma_{dust}$, Equation~\ref{e_dustmass} is first fitted to the 250 and 350~$\mu$m data points.  We then compare the 160~$\mu$m surface brightness from the fitted SED to the measured value, and if the fitted value is higher, we recalculate the dust surface density using the 160-350~$\mu$m data.  We then use the 100~$\mu$m and then 70~$\mu$m in the same way to constrain the best fitting dust SED.  Uncertainties are determined using a monte carlo approach based on background noise measurements.  In Appendix~\ref{a_dustmass}, we used a broad variety of models with multiple thermal components to test this approach to estimating $\sigma_{dust}$.  We found that the values derived using our approach will usually fall within 20\% and very often within 10\% of the input model values for a wide variety of scenarios.

\section{Relation of the PAH/dust ratios to stellar populations}

The PAH/dust ratio ($I_{PAH}/\sigma_{dust}$) does not vary in the same way among all galaxies, which is consistent with the differing recently-published results on PAH excitation \citep{calapa14, lu14, jones15}.  The best way to present these results is to separate the sample galaxies into separate subgroups that exhibit similar phenomenology in terms of their PAH/dust variations.  These subsets are discussed in separate sections below.

\subsection{Galaxies where the PAH/dust ratios correlate directly with star formation tracers}
\label{s_comprat_sfr}

\begin{table*}
\centering
\begin{minipage}{114mm}
\caption{Statistics for relation of log($I_{PAH}/\sigma_{dust}$) to other quantities where PAH excitation is directly linked to star formation}
\label{t_comprat_sfr}
\begin{tabular}{@{}lccccc@{}}
\hline
Galaxy &
  Number &
  \multicolumn{4}{c}{Correlation Coefficient for Relations of Quantities to log($I_{PAH}/\sigma_{dust}$)}\\
&
  of Bins &
  log($I_{154 nm}$) &
  log($I_{154nm + 24\mu m}$) &
  log($I_{24 \mu m}$) &
  log($I_{3.6 \mu m}$) \\
\hline
NGC 628  &           225 &     0.77 &     0.88 &     0.91 &     0.83 \\
NGC 3184 &           130 &     0.74 &     0.82 &     0.84 &     0.77 \\
NGC 3938 &            74 &     0.91 &     0.91 &     0.90 &     0.85 \\
NGC 4254 &           107 &     0.91 &     0.90 &     0.89 &     0.82 \\
NGC 4303 &            96 &     0.94 &     0.94 &     0.93 &     0.85 \\
NGC 4321 &           148 &     0.90 &     0.90 &     0.89 &     0.85 \\
NGC 4548 &            69 &     0.37 &     0.76 &     0.76 &     0.70 \\
NGC 4579 &            63 &     0.87 &     0.96 &     0.95 &     0.88 \\
NGC 4725 &           157 &     0.63 &     0.83 &     0.88 &     0.63 \\
NGC 5055 &           255 &     0.95 &     0.96 &     0.96 &     0.89 \\
NGC 5248 &            66 &     0.89 &     0.90 &     0.89 &     0.81 \\
\hline
\end{tabular}
\end{minipage}
\end{table*}

For 11 of the 25 galaxies in our sample, the variations in the PAH/dust ratio are associated with star formation in a simple, direct way.  Images of the PAH/dust ratio compared to 154~nm, 3.6~$\mu$m, and 24~$\mu$m emission  are shown in Figure~\ref{f_map_sfr} for two of the galaxies in this subset.  The structures in the PAH/dust ratio maps mirror the star forming structures seen in the far-ultraviolet or mid-infrared images, particularly along spiral and ring structures.  In contrast, the PAH/dust ratio is not as strong in regions where the 3.6~$\mu$m is relatively high compared to star formation, as can be seen particularly well in the centre of NGC~628.  It could be argued that the PAH/dust ratio varies in part because of variations in the dust surface density relative to the number of photons from the stellar population that excites the PAHs, but NGC~628 is a clear example of where that is not happening.  The dust surface density is almost the same in the centre and northern arm of that galaxy, yet the PAH/dust ratio is significantly higher where star formation is stronger in the northern arm.

The relations of the PAH/dust ratios to the 154~nm, 154~nm + 24~$\mu$m, 24~$\mu$m, and 3.6~$\mu$m emission for the measurements in 24~arcsec bins are shown in Figure~\ref{f_comprat_sfr}.  Weighted Pearson correlation coefficients\footnote{Weighted Pearson correlation coefficients are used throughout the paper.  By using weights based on the uncertainties in the logarithms of the PAH/dust ratios, we are able to compensate for data that show scatter because of their low signal-to-noise levels.} for these relations are listed in Table~\ref{t_comprat_sfr}.  If we look at the rows in either the figure or the table, we see that, for each galaxy the correlation coefficients for at least two relations involving star formation metrics are $>$0.05 higher than the coefficients for the relations with the 3.6~$\mu$m emission.

The PAH/dust ratio does not always clearly correlate with one star formation tracer better than the others, so it is difficult to conclude that PAHs are excited either primarily by the unobscured ultraviolet photons escaping star forming region or by the photons that also heat dust locally around star forming regions.  However, in five galaxies, the relations of the PAH/dust ratios to 154~nm emission have notably lower correlation coefficients than the relations between the ratios and either of the other star formation metrics.  Moreover, the 3.6~$\mu$m emission correlates with the PAH/dust ratios better than the 154~nm emission does in some galaxies.

One of the notable features of the trends in Figure~\ref{f_comprat_sfr} is how the relations of the PAH/dust ratio versus 3.6~$\mu$m emission tend to droop at the high surface brightness ends of the relations.  This is most easily seen in the data for NGC~628, NGC~4254, NGC~4303, NGC~4321, NGC~4725, and NGC~5055.  We generally interpret this as indicative of how PAH excitation is not dependent on 3.6 micron surface brightness.  If PAH excitation is linked to star formation and if star formation activity is low in the centres of these galaxies where the 3.6~$\mu$m emission is strongest, then the relations should bend.  These bends could also be interpreted as places where the PAH/dust ratios are low because the dust surface densities are high.  This could also explain why the relations of the PAH/dust ratios versus 24~$\mu$m emission also dips in the centres of NGC~3184 and NGC~4254.  However, even when the affected bins from the central regions are excluded from the analysis, the PAH/dust ratios still generally correlate better with one or more star formation metrics for this subsample of galaxies.  Also, NGC~4548, NGC~4579, and NGC~4725 are specific cases where the dust surface density peaks in spiral arm structures outside the centre of the galaxies but the PAH/dust ratios appear suppressed in the galaxies' centres.  Moreover, NGC~628 as discussed above is a clear example where the dust density variations do not appear strongly linked to variations in the PAH/dust ratio.

Seven of the galaxies listed in Table~\ref{t_comprat_sfr} contain AGN.  We performed tests where we removed the central 3$\times$3 bins to see if excluding the AGN affected our analysis.  For six of the galaxies, we still find that the PAH/dust ratios are more strongly correlated to one or more of the star formation tracers than to 3.6~$\mu$m emission.  The exception is NGC 3938, where the correlation coefficients for all relations fall within the range 0.89-0.92 if we exclude the central region.  This is in part because neither the star formation tracers nor the PAH/dust ratio peak in the centre of NGC~3938 but the 3.6~$\mu$m does peak in the centre.  The PAH/dust ratio could be low in the nucleus of the galaxy because the ratio is not connected to the evolved stars producing the 3.6~$\mu$m emission or because the environment around the AGN affects the PAH emission.  Given the lack of robustness in the analysis of NGC~3938, the results from this specific galaxy should be treated with caution.

\subsection{Galaxies where the PAH/dust ratios correlate with star formation in complex ways}
\label{s_comprat_sfrcomplex}

\begin{figure*}
\epsfig{file=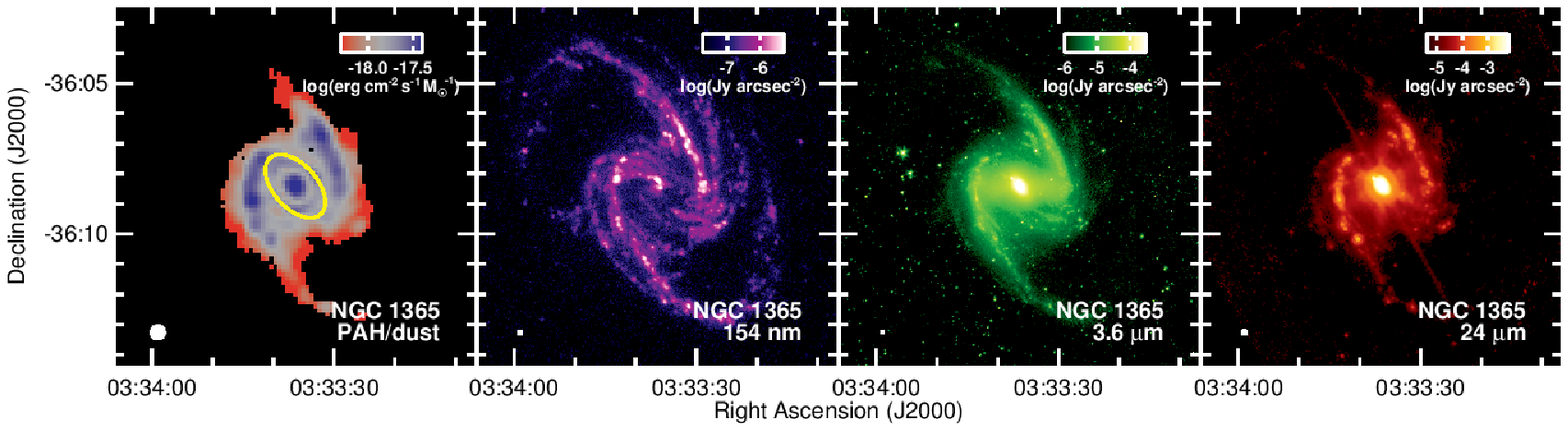}

\vspace{0.3cm}

\epsfig{file=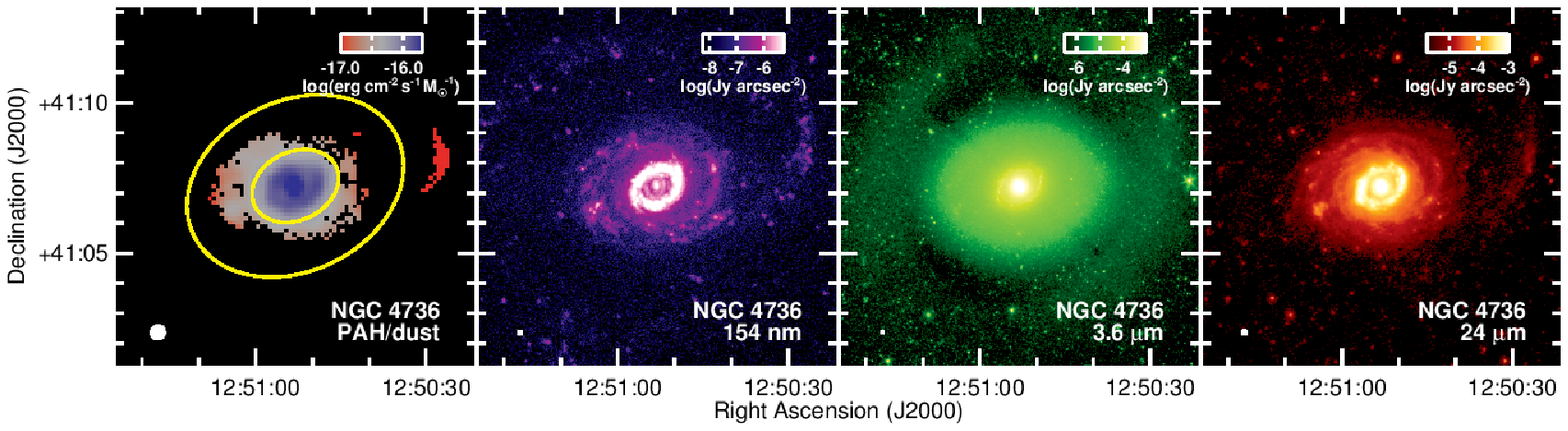}

\vspace{0.3cm}

\epsfig{file=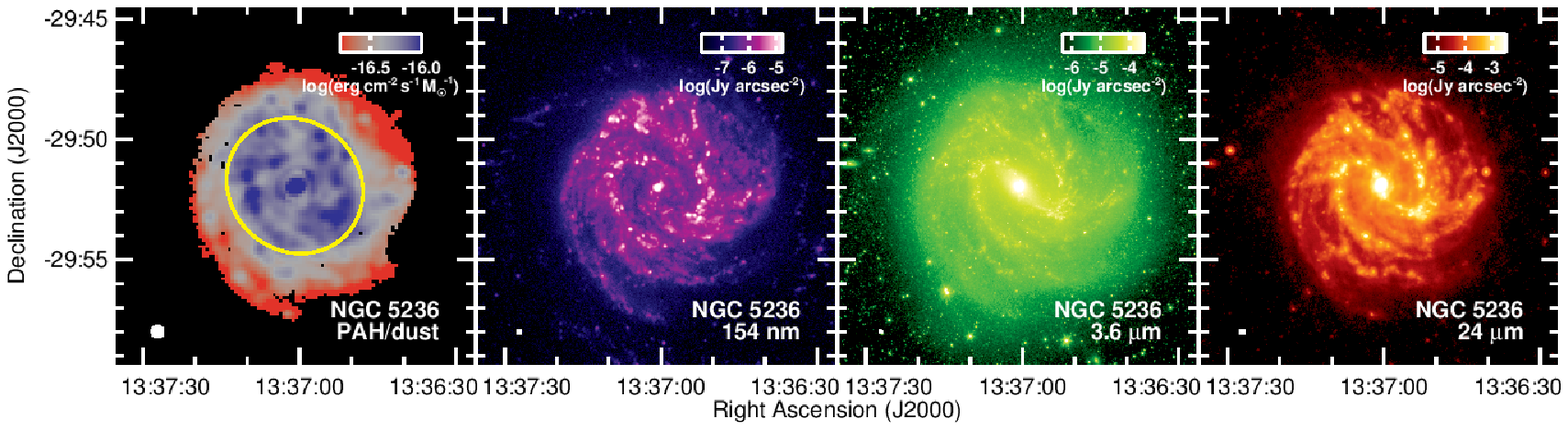}

\vspace{0.3cm}

\epsfig{file=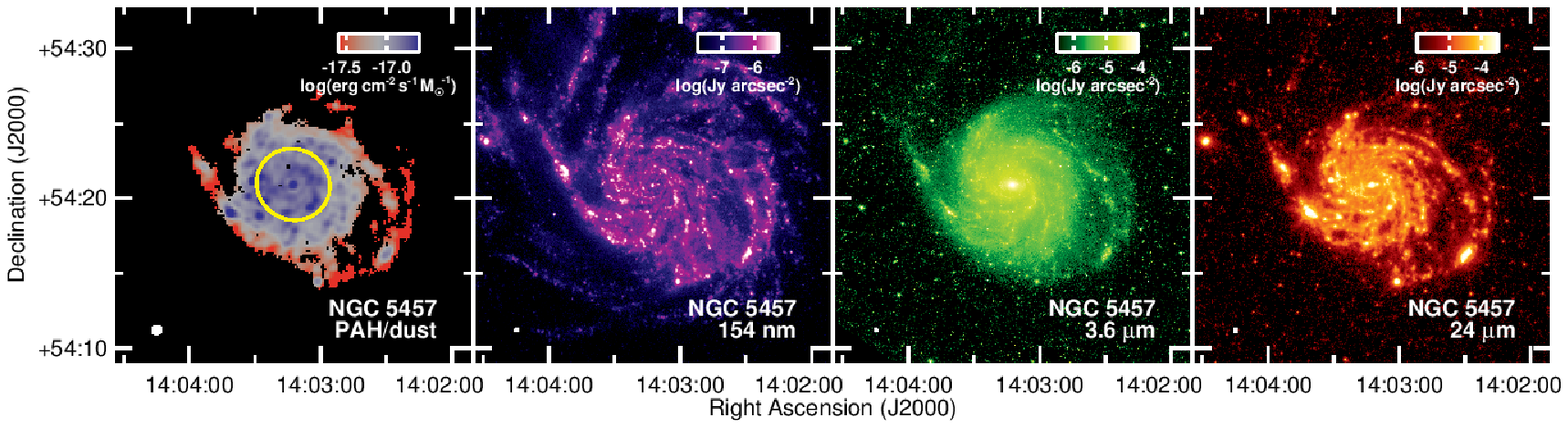}

\caption{Maps of the PAH/dust ratio, 154~nm emission, 3.6~$\mu$m emission, and 24~$\mu$m emission for four galaxies where the PAH excitation is linked to the star formation in only part of the disc.  The ellipses in the PAH/dust ratio maps indicate radii where the relations between the ratios and star formation tracers change; the corresponding radii are indicated in Table~\ref{f_comprat_sfrcomplex}.  The image format is otherwise the same as for Figure~\ref{f_map_sfr}; see the caption of that figure for additional details.}
\label{f_map_sfrcomplex}
\end{figure*}

\begin{figure*}
\epsfig{file=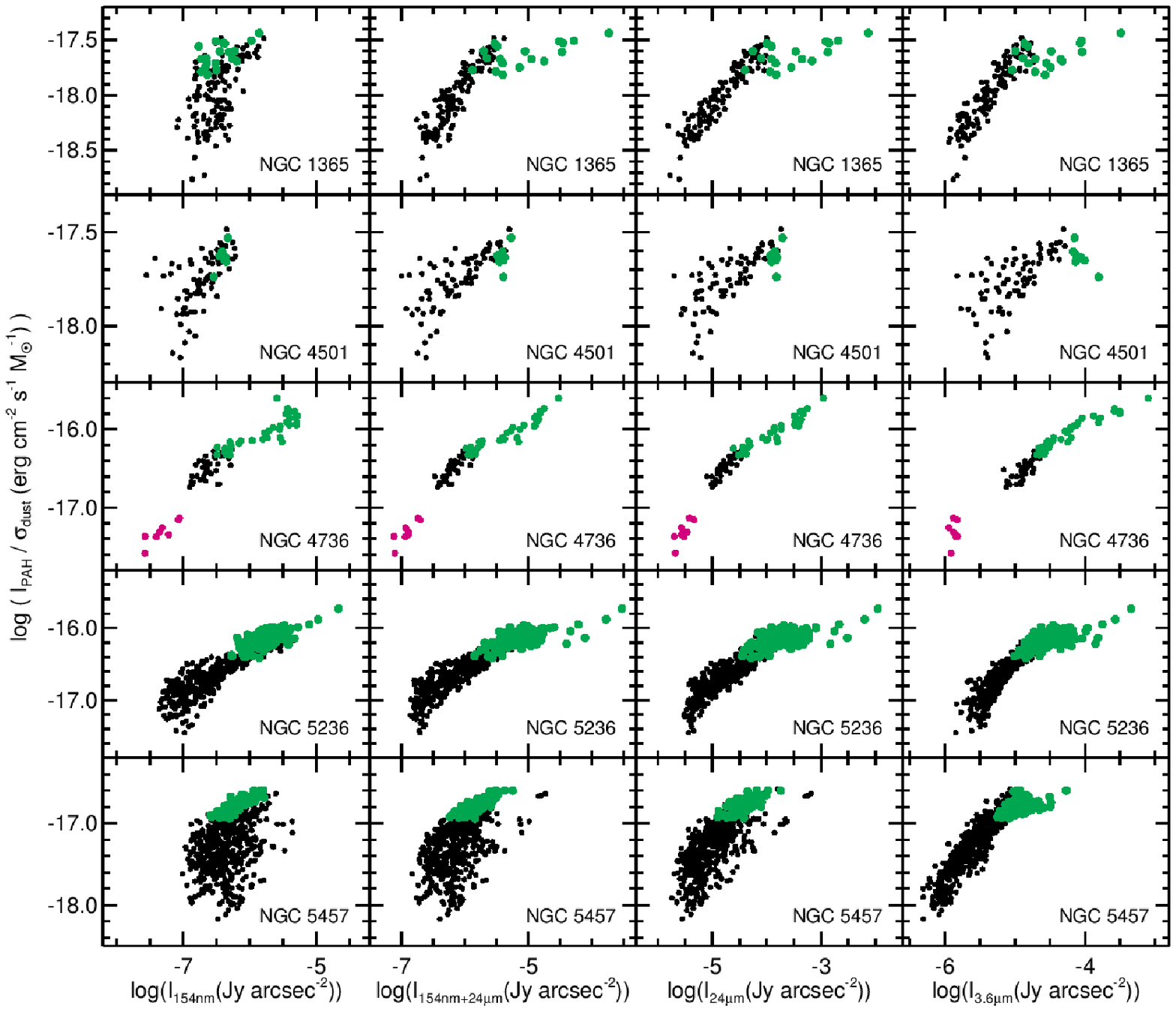}
\caption{Plots of the PAH/dust ratio versus other quantities as measured within 24~arcsec square bins for the galaxies where the ratios are correlated with one or more star formation tracers in only part of the galaxies' discs.  The data points coloured green are from inner radii at the locations specified in Table~\ref{t_comprat_sfrcomplex}.  The magenta points for NGC~4736 are at locations where $r$$>$5~kpc.  The image format is otherwise the same as for Figure~\ref{f_comprat_sfr}; see the caption of that figure for additional details.  Statistics for the relations are in Table~\ref{t_comprat_sfrcomplex}.}
\label{f_comprat_sfrcomplex}
\end{figure*}

\begin{table*}
\centering
\begin{minipage}{135mm}
\caption{Statistics for relation of log($I_{PAH}/\sigma_{dust}$) to other quantities where PAH excitation is linked to star formation in complex ways}
\label{t_comprat_sfrcomplex}
\begin{tabular}{@{}lccccc@{}}
\hline
Galaxy &
  Number &
  \multicolumn{4}{c}{Correlation Coefficient for Relations of Quantities to log($I_{PAH}/\sigma_{dust}$)}\\
&
  of Bins &
  log($I_{154 nm}$) &
  log($I_{154nm + 24\mu m}$) &
  log($I_{24 \mu m}$) &
  log($I_{3.6 \mu m}$) \\
\hline
NGC 1365 &           145 &     0.64 &     0.83 &     0.85 &     0.84 \\
NGC 1365 ($r$$\leq$6.5~kpc) &
                      19 &     0.81 &     0.90 &     0.90 &     0.87 \\
NGC 1365 ($r$$>$6.5~kpc) &
                     126 &     0.54 &     0.92 &     0.94 &     0.89 \\
NGC 4501 &            82 &     0.74 &     0.78 &     0.78 &     0.58 \\
NGC 4501 ($r$$>$3~kpc) &            
                      75 &     0.75 &     0.87 &     0.86 &     0.82 \\
NGC 4736 &            79 &     0.88 &     0.97 &     0.97 &     0.96 \\
NGC 4736 ($r$$<$2~kpc) &
                      35 &     0.65 &     0.95 &     0.96 &     0.96 \\
NGC 4736 (2~kpc$\leq$$r$$\leq$5~kpc) &
                      36 &     0.62 &     0.82 &     0.87 &     0.81 \\
NGC 5236 &           544 &     0.91 &     0.90 &     0.89 &     0.88 \\
NGC 5236 ($r$$\leq$4~kpc) &
                     159 &     0.89 &     0.86 &     0.85 &     0.83 \\
NGC 5236 ($r$$>$4~kpc) &
                     385 &     0.88 &     0.90 &     0.88 &     0.90 \\
NGC 5457 &           657 &     0.46 &     0.67 &     0.78 &     0.88 \\
NGC 5457 ($r$$\leq$5~kpc) &
                     112 &     0.86 &     0.85 &     0.75 &     0.46 \\
NGC 5457 ($r$$>$5~kpc)&
                     545 &     0.44 &     0.60 &     0.73 &     0.92 \\
\hline
\end{tabular}
\end{minipage}
\end{table*}

Five of the sample galaxies show some connection between the PAH/dust ratios and star formation, but the relations do not show the type of direct relation seen in the galaxies discussed in Section~\ref{s_comprat_sfr}.  Instead, the correlation between the PAH/dust ratios and one or more star formation tracers for each galaxy is seen only in part of the disc.  Images of three of the galaxies are shown in Figure~\ref{f_map_sfrcomplex}.  The relations between the PAH/dust ratios and other quantities measured within the 24~arcsec binned regions are shown in Figure~\ref{f_comprat_sfrcomplex}, with the statistics on these relations listed in Table~\ref{t_comprat_sfrcomplex}. 

NGC~1365 and NGC~4501 are galaxies where the PAH/dust ratios are more strongly correlated with one or more star formation tracers than with 3.6~$\mu$m emission in the outer discs but where the relation dips in the centres of the galaxies, as seen very clearly in the plots of the binned data in Figure~\ref{f_comprat_sfrcomplex}.  A few other galaxies discussed in the previous section may also exhibit such phenomena, particularly NGC~3938 and NGC~4254, but the dip is seen only within the central bin and is not as large in magnitude.

The radius at which this occurs is different for each galaxy.  In NGC~1365, data from radii less than 6.5~kpc, which includes the central starburst and a significant portion of the bar, drop below the relation.  In this region the PAH/dust ratios correlated equally well with 24~$\mu$m, 154~nm + 24~$\mu$m, and 3.6~$\mu$m emission.  In NGC~4501, data from radii less than 3~kpc do not follow the relation, but since we only had 7 data points covering this region, we did not attempt to perform any analysis on these data.

The reason why the PAH/dust ratios drop in the centres of these  two galaxies is not entirely clear.  The nucleus of NGC~1365 is sufficiently bright at 8~$\mu$m that it may have saturated the detector, which might explain the observed dip in the PAH/dust ratio for the centre of this galaxy but possibly not the more extended region around it.  Moreover the centre of NGC~4501 is similar in brightness to the centres of other galaxies, so saturation effects could not explain the dip in the PAH/dust ratios in its centre.  NGC~1365 has an infrared-bright composite AGN/starburst nucleus, and NGC~4501 also has a Seyfert nucleus, and since PAH emission from AGN appears suppressed relative to dust continuum emission, the AGN could be responsible for the dips in the centres of these galaxies.  However, many of the other sample galaxies have AGN, and the PAH/dust ratios do not dip in the centres of these galaxies like in NGC~1365 and NGC~4501.  Having said that, the AGN could still be factors in suppressing PAH/dust ratios in these two galaxies given that we have no other plausible explanation for the phenomenon.  Alternately, the centres of these two galaxies could be cases where the dust surface densities are very high relative to the number of PAH-exciting photons, thus making the PAH/dust ratios appear uncorrelated with any stellar population.

\begin{figure*}
\epsfig{file=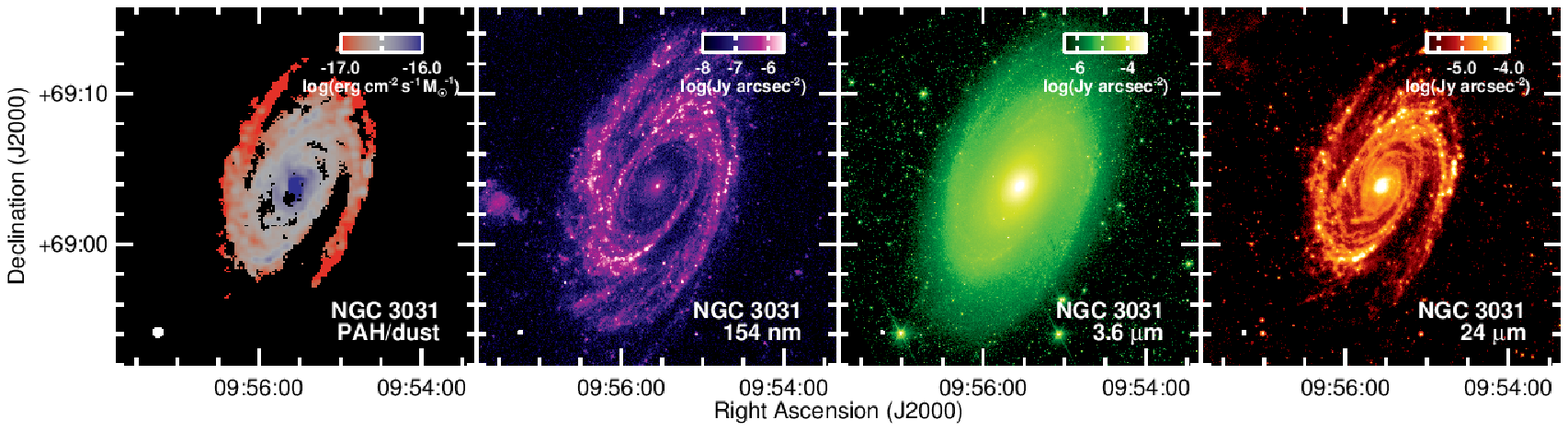}

\vspace{0.3cm}

\epsfig{file=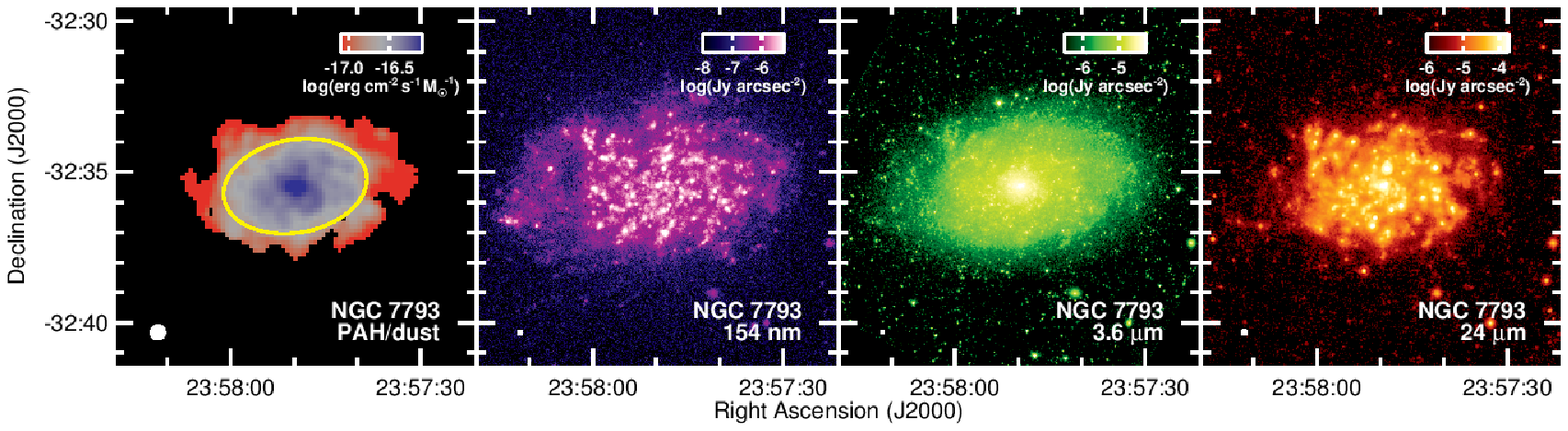}
\caption{Maps of the PAH/dust ratio, 154~nm emission, 3.6~$\mu$m emission, and 24~$\mu$m emission for two example galaxies where the PAH excitation is linked to the evolved stellar population (as traced by the 3.6~$\mu$m emission).  The ellipse in the PAH/dust ratio map of NGC~7793 indicates the region within which the PAH/dust ratios correlate best with the 3.6~$\mu$m emission.  The image format is otherwise the same as for Figure~\ref{f_map_sfr}; see the caption of that figure for additional details.}
\label{f_map_evolved}
\end{figure*}

\begin{figure*}
\epsfig{file=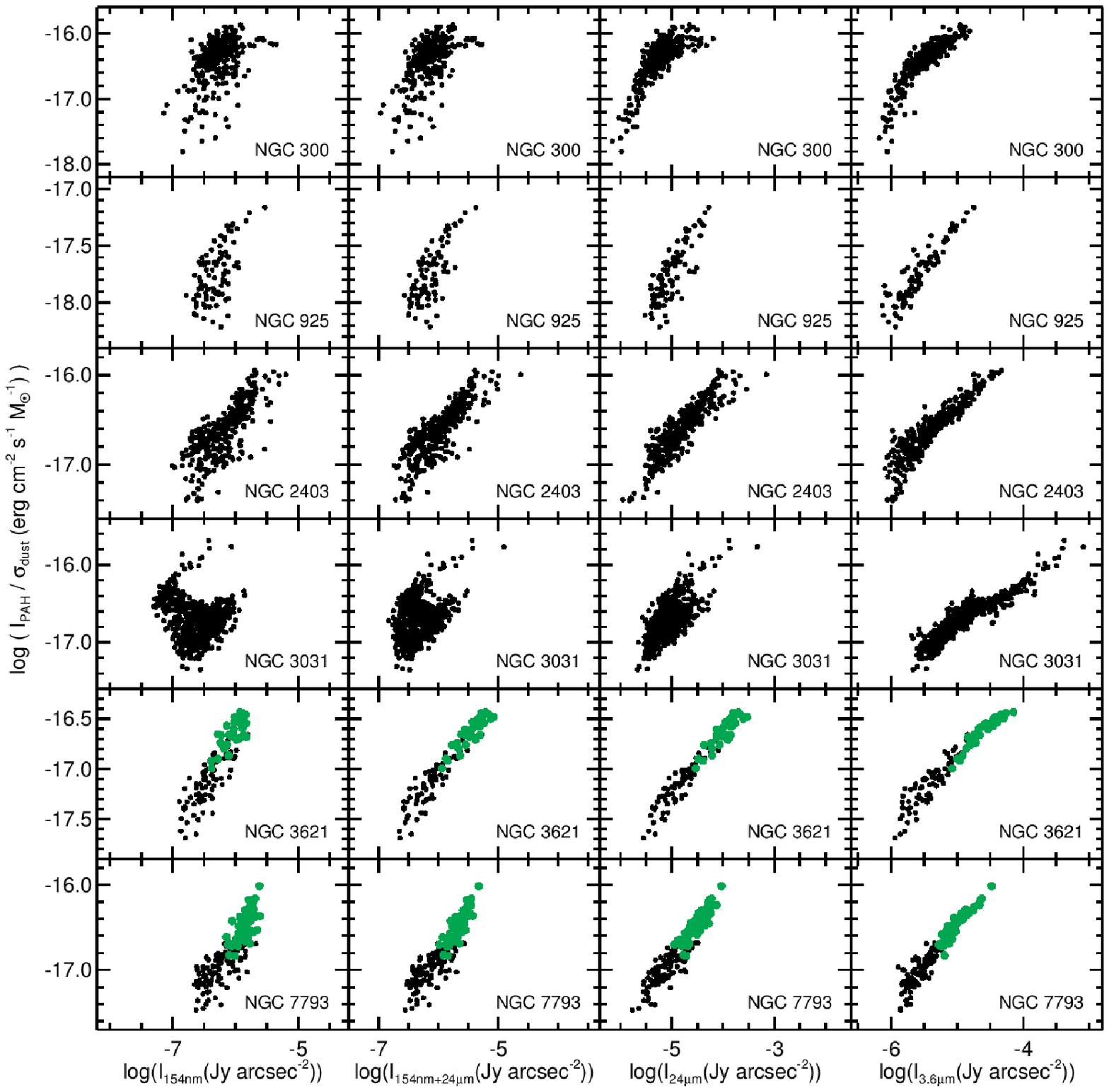}
\caption{Plots of the PAH/dust ratio versus other quantities as measured within 24~arcsec square bins for the galaxies where the ratios are correlated most strongly with 3.6~$\mu$m emission.  The data points coloured green are for locations with $r$$\le$4~kpc in NGC~3621 and $r$$\le$2.5~kpc in NGC~7793 where the PAH/dust ratios correlate the best with 3.6~$\mu$m emission.  The image format is otherwise the same as for Figure~\ref{f_comprat_sfr}; see the caption of that figure for additional details.  Statistics for the relations are in Table~\ref{t_comprat_evolved}.}
\label{f_comprat_evolved}
\end{figure*}

NGC~4736 has an AGN, an infrared-bright circumnuclear star forming ring, and an outer ring structure.  The data for the galaxy overall do not show a definitive difference between how the PAH/dust ratios correlate with either star formation tracers or 3.6~$\mu$m emission as seen in both the plots of the binned data in Figure~\ref{f_comprat_sfrcomplex} or the correlation coefficients listed in Table~\ref{t_comprat_sfrcomplex}, but this could be because of the strong contrast in all bands between the inner and outer regions.  We therefore divided the binned data for this galaxy into three subsets: data sampling the circumnuclear ring at $r\leq$2~kpc, data sampling the middle of the galaxy disc at 2$<r\leq$5~kpc, and data sampling the outer ring at $r>$5~kpc.  

The inner and middle regions yield different results.  The inner region still shows no significant differences in the correlations of the PAH/dust ratio to either 24~$\mu$m, 154~nm + 24~$\mu$m, or 3.6~$\mu$m emission.  The middle of the disc shows that the PAH/dust ratios are slightly better correlated with 24~$\mu$m than with 3.6~$\mu$m emission, although it is difficult to see any star forming structures in the maps in Figure~\ref{f_map_sfrcomplex} that illustrate this in a clear, qualitative way.  Additionally, the 5.8 and 8.0~$\mu$m images are affected by muxbleed, and even though we have tried to exclude the 24~arcsec bins most strongly affected by this, it could still affect the results.  Only 8 data points are available for the outer dust ring, so it is difficult to draw any secure conclusions from those data.  None the less, these results indicate that the PAH/dust ratios may be linked to star formation within part of the disc of NGC~4736.

\begin{table*}
\centering
\begin{minipage}{140mm}
\caption{Statistics for relation of log($I(\mbox{PAH})/\sigma_{dust})$) to other quantities where PAH excitation is linked to evolved stellar populations}
\label{t_comprat_evolved}
\begin{tabular}{@{}lccccc@{}}
\hline
Galaxy &
  Number &
  \multicolumn{4}{c}{Correlation Coefficient for Relations of Quantities to log($I_{PAH}/\sigma_{dust}$)}\\
&
  of Bins &
  log($I_{154 nm}$) &
  log($I_{154nm + 24\mu m}$) &
  log($I_{24 \mu m}$) &
  log($I_{3.6 \mu m}$) \\
\hline
NGC 300  &           341 &     0.49 &     0.56 &     0.71 &     0.86 \\
NGC 925  &            89 &     0.69 &     0.80 &     0.86 &     0.94 \\
NGC 2403 &           339 &     0.81 &     0.85 &     0.88 &     0.95 \\
NGC 3031 &           654 &    -0.11 &     0.40 &     0.68 &     0.92 \\
NGC 3621 &           104 &     0.87 &     0.95 &     0.96 &     0.97 \\
NGC 3621 ($r$$\le$4~kpc) &
                      33 &     0.63 &     0.83 &     0.82 &     0.94 \\
NGC 3621 ($r$$>$4~kpc) &
                      71 &     0.81 &     0.91 &     0.94 &     0.93 \\
NGC 7793 &           179 &     0.86 &     0.91 &     0.94 &     0.97 \\
NGC 7793 ($r$$\le$2.5~kpc) &
                      69 &     0.70 &     0.86 &     0.89 &     0.95 \\
NGC 7793 ($r$$>$2.5~kpc) &
                     110 &     0.70 &     0.76 &     0.82 &     0.85 \\
\hline
\end{tabular}
\end{minipage}
\end{table*}

In NGC~5236, the PAH/dust ratios clearly trace the spiral arms, which at first would imply that the PAHs are excited by star forming regions within the arms, but the spiral arms are also prominent sources of 3.6~$\mu$m emission.  Consequently, the PAH/dust ratios from the binned data correlate equally well with the 154~nm, 24~$\mu$m, and 3.6~$\mu$m emission.  However, PAH/dust ratios within radii of 4~kpc correlate better with 154~nm emission than with the other quantities.  In particular, the correlation coefficient for the relation with the 154~nm emission is 0.06 higher than the one with the 3.6~$\mu$m emission.  This region includes the initial bends in the spiral arms at the ends of the weak bar, and significant offsets are seen between the ultraviolet and infrared emission in these bends \citep{bendo15, jones15}.  As also identified by \citet{jones15}, the PAHs appear to be excited most strongly in these offset regions.  In the outer part of the disc, though, the differences between the ultraviolet, near-infrared, and mid-infrared structures may not be significant enough at these angular resolutions to see differences between the structures seen in the different bands or to show that one band correlates more strongly with the PAH/dust ratios than the others.

NGC~5457 appears to be a galaxy where the stellar population exciting the PAH emission differs between the inner and outer parts of the galaxy.  The maps of the PAH/dust ratio in Figure~\ref{f_map_sfrcomplex} show that PAH emission is locally enhanced near star forming regions.  In the centre of the galaxy, the ratios do not appear smooth and do not peak in the centre like the 3.6~$\mu$m emission does.  However, the brightest star forming regions in far-ultraviolet and mid-infrared emission, which are located in the outer parts of the disc, correspond to locations with relatively modest enhancements in PAH emission.  Confusingly, the statistics in Table~\ref{t_comprat_sfrcomplex} for the binned data covering the entire optical disc of NGC~5457 indicate that the PAH/dust ratios correlate best with the 3.6~$\mu$m emission.  The plots of the binned data in Figure~\ref{f_comprat_sfrcomplex} show that the PAH/dust ratios exhibit a substantial amount of scatter when plotted versus any star formation tracer but less scatter when plotted versus 3.6~$\mu$m emission, although the ratios flatten when $I_\nu$(3.6~$\mu$m)$>$$10^{-5}$ Jy arcsec$^{-2}$.

When NGC 5457 is subdivided into inner and outer regions, the results look different.  The PAH/dust ratios for the inner region appear very strongly correlated with far-ultraviolet emission.  For the statistical analysis, we chose data with $r\leq$5~kpc, which is the approximate enclosing radius where the correlation coefficient for the relation between the PAH/dust ratio and the 154~nm emission reaches a maximum.  At $r>$5~kpc, the correlation coefficient for the relation between the ratio and the 3.6~$\mu$m emission is much stronger.  The PAH emission in the outer disc still appears enhanced in locations with strong star formation activity, but so does the 3.6~$\mu$m emission.  This suggests a scenario also discussed in Section~\ref{s_comprat_evolved} where either young ultraviolet-luminous stars partially contribute to PAH excitation and 3.6~$\mu$m emission or where PAH excitation is enhanced by light from massive red stars that recently evolved from the ultraviolet-luminous stars.  None the less, these results show that the PAH excitation is more strongly linked to unobscured ultraviolet light from young stars in the centre of NGC~5457 but is more strongly linked to evolved stars in the outer part of the galaxy.

\subsection{Galaxies where the PAH/dust ratios correlate with the evolved stellar populations}
\label{s_comprat_evolved}

In another six of the 25 galaxies, the PAH/250~$\mu$m ratios are more strongly correlated with 3.6~$\mu$m emission than any of the star formation metrics.  Figure~\ref{f_map_evolved} shows maps of NGC 3031 and NGC~7793 as examples of these galaxies.  The PAH/dust ratio maps have smooth gradients with radius that are similar to the gradients in the 3.6~$\mu$m emission.  Relations for the 24~arcsec binned data are shown for all six galaxies in Figure~\ref{f_comprat_evolved}, with the statistics of the relations given in Table~\ref{t_comprat_evolved}.  When looking at the rows of data for NGC~300, NGC~925, NGC~2403, or NGC~3031, we can see that the relation between the PAH/dust ratio and 3.6~$\mu$m emission has a correlation coefficient that is at least 0.05 higher than the relations involving star formation metrics in each of these galaxies.  In NGC~3621 and NGC~7793, the results for all of the data within each galaxy is more ambiguous, but when only the data from central regions are used, the PAH/dust ratios appear much more clearly correlated with 3.6~$\mu$m emission.  This is most likely because PAHs are predominantly excited by evolved stars only where the surface brightnesses of these stars are higher.  Four of these galaxies have nuclei powered by star formation, but NGC~3031 and NGC~3621 contain AGN.  However, the relative differences between the correlation coefficients do not change significantly if we exclude the central 3$\times$3 bins in these two galaxies.

In the maps of these galaxies, including the two shown in Figure~\ref{f_map_evolved}, the PAH/dust ratios sometimes appear slightly enhanced in locations with relatively strong star formation activity.  This could be caused by one of two phenomena.  First, it is possible that the enhancement in star formation activity leads to local enhancements in PAH excitation.  This type of scenario could explain why the analysis of the binned data yield more ambiguous results for the outer regions of NGC~3621 and NGC~7793 where the brightness of the evolved stellar discs becomes very low.  Second, it is possible that the ultraviolet-luminous stars themselves do not significantly excite the PAHs (or that they tend to destroy PAHs locally) but that the red supergiants that form soon after the onset of star formation both excite PAHs and enhance emission in the 3.6~$\mu$m band.  This would preserve the correlation between the PAH/dust ratio and 3.6~$\mu$m emission.  

\begin{figure*}
\epsfig{file=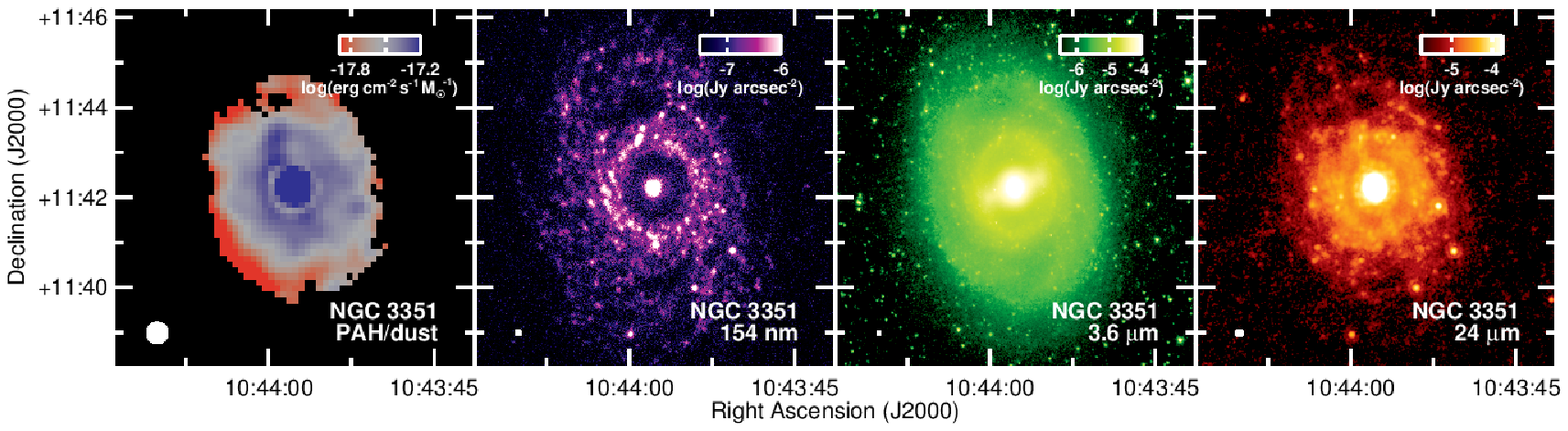}

\vspace{0.3cm}

\epsfig{file=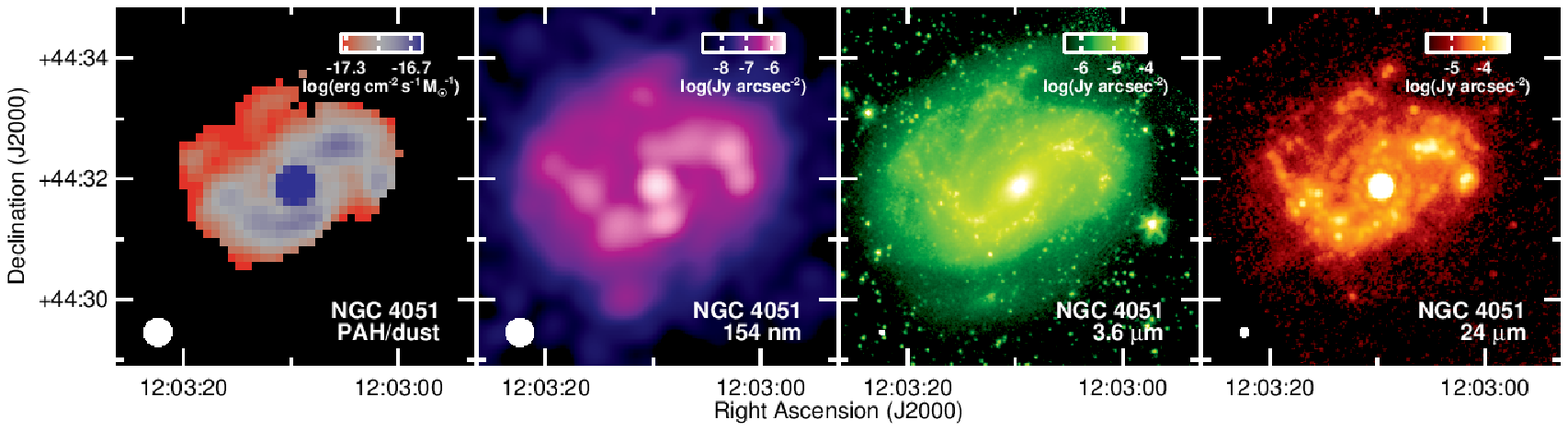}
\caption{Maps of the PAH/dust ratio, 154~nm emission, 3.6~$\mu$m emission, and 24~$\mu$m emission for two galaxies where the analysis yields ambiguous results regarding how PAHs are excited.   The version of the NGC~4051 154~nm image shown here has been convolved with kernels from \citet{aniano11} to match the PSF to that of the 350~$\mu$m data, because the far-ultraviolet observations of this galaxy were very shallow and relatively noisy.  The image format is otherwise the same as for Figure~\ref{f_map_sfr}; see the caption of that figure for additional details.}
\label{f_map_ambig}
\end{figure*}

\begin{figure*}
\epsfig{file=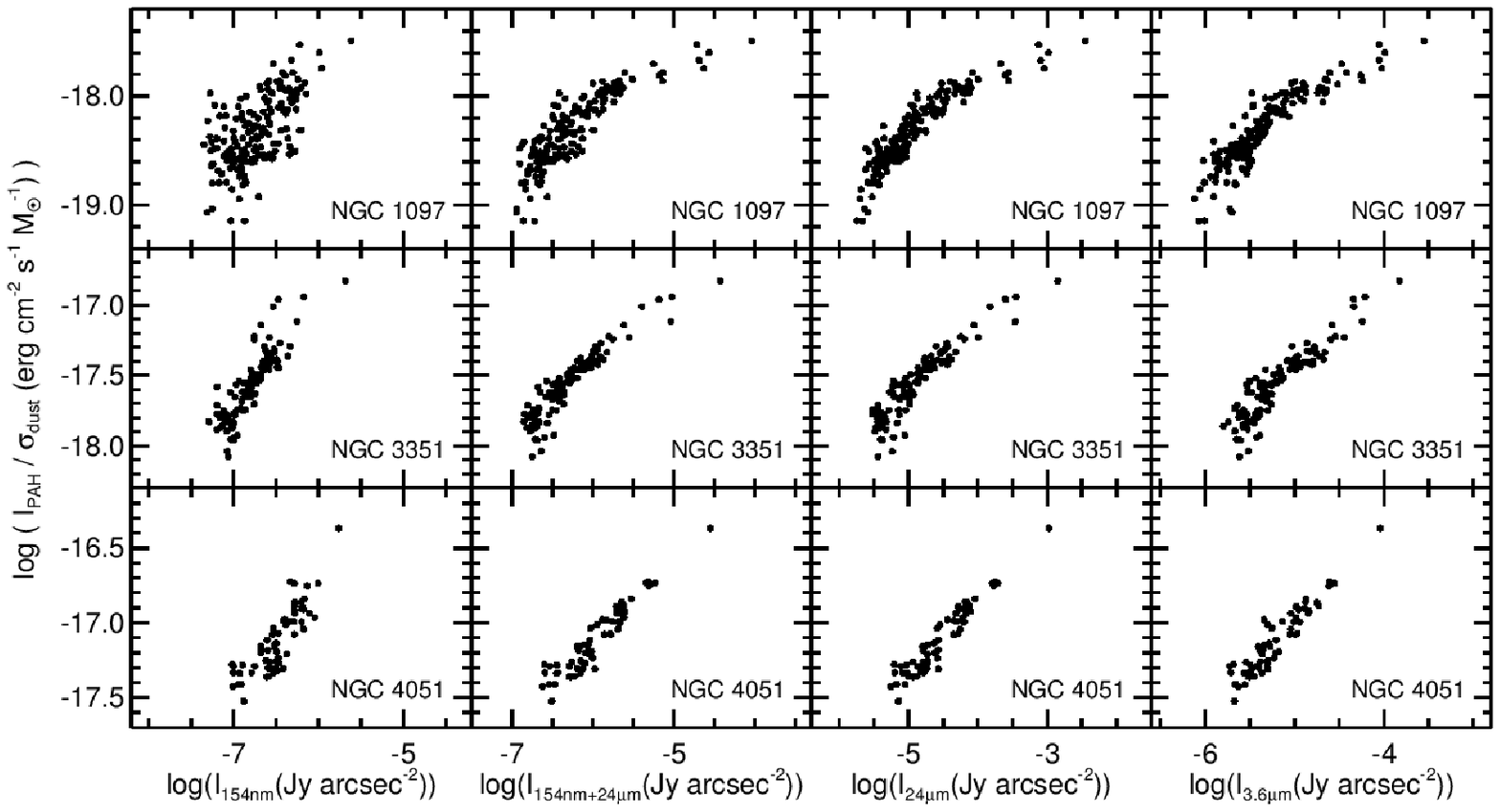}
\caption{Plots of the PAH/dust ratios versus other quantities as measured within 24~arcsec square bins for the galaxies where the analysis yields ambiguous results regarding how PAHs are excited.  The image format is the same as for Figure~\ref{f_comprat_sfr}; see the caption of that figure for additional details.  Statistics for the relations are in Table~\ref{t_comprat_ambig}.}
\label{f_comprat_ambig}
\end{figure*}

\begin{table*}
\centering
\begin{minipage}{114mm}
\caption{Statistics for relation of log($I_{PAH}/\sigma_{dust}$) to other quantities where the results are ambiguous}
\label{t_comprat_ambig}
\begin{tabular}{@{}lccccc@{}}
\hline
Galaxy &
  Number &
  \multicolumn{4}{c}{Correlation Coefficient for Relations of Quantities to log($I_{PAH}/\sigma_{dust}$)}\\
&
  of Bins &
  log($I_{154 nm}$) &
  log($I_{154nm + 24\mu m}$) &
  log($I_{24 \mu m}$) &
  log($I_{3.6 \mu m}$) \\
\hline
NGC 1097 &           175 &     0.85 &     0.93 &     0.95 &     0.95 \\
NGC 3351 &           107 &     0.92 &     0.96 &     0.96 &     0.96 \\
NGC 4051 &            61 &     0.86 &     0.96 &     0.97 &     0.96 \\
\hline
\end{tabular}
\end{minipage}
\end{table*}

\subsection{Galaxies where the PAH/dust ratios yield ambiguous results}
\label{s_comprat_ambig}

The remaining three sample galaxies all produce results that are qualitatively and quantitatively ambiguous.  The maps of the PAH/dust ratio shown for NGC~3351 and NGC~4051 in Figure~\ref{f_map_ambig} look very similar to the maps for both the 3.6 and 24~$\mu$m emission.  The 24~arcsec binned data plotted in Figure~\ref{f_comprat_ambig} do not show any visually apparent difference in the trends for the PAH/dust ratios plotted versus 154~nm + 24~$\mu$m, 24~$\mu$m, or 3.6~$\mu$m emission.  The correlation coefficients for the relations for each galaxy as listed in each row of Table~\ref{t_comprat_ambig} are all similar.  

These results arise at least in part because both the 3.6 and 24~$\mu$m bands trace very similar structures within these galaxies as seen at angular resolutions of 25~arcsec, so any quantity that correlates with one of the bands also correlates with the other.  Two of the galaxies have circumnuclear star forming rings that appear bright in all infrared bands, and an infrared-bright AGN lies at the centre of the ring in NGC~1097.  This helps to generate the similarities among the relations seen in Figure~\ref{f_comprat_ambig}, but even when data from the central regions are excluded, the relations of PAH/dust ratio to either 3.6 or 24~$\mu$m emission still look similar.  NGC~4051 has an AGN that appears bright in all the infrared bands as well, but none of the relations in Figure~\ref{f_comprat_ambig} change significantly when the AGN is included or excluded.  The 8~$\mu$m images of two galaxies are also affected by muxbleed (despite our efforts to remove the artefacts and despite masking any obviously-affected 24~arcsec bins), which complicates or analysis.

Analyses on subregions of the galaxies do not yield any significant results.  It is possible to select areas in the outer disc of NGC~3351 where the PAH/dust ratios show a slightly stronger correlation with a star formation tracer than 3.6~$\mu$m emission, but this result is sensitive to the radius used for selecting data points.  

With these three galaxies, it is impossible to state just from this analysis exactly how the PAHs are excited.  It is possible that just young stars in star forming regions, just evolved stars, or a combination of these stellar populations are responsible for exciting PAHs.  Either a different analytical approach or higher angular resolution observations would be needed to determine how the PAHs are excited.

\section{Comparison of PAH/dust ratio gradients to log(O/H) gradients}
\label{s_radialvar}

Results from {\it Spitzer} demonstrated that PAH emission appeared strongly suppressed compared to dust emission in other bands in dwarf galaxies and other environments with low gas phase oxygen abundances (\citealt{engelbracht05, engelbracht08, madden06, calzetti07, gordon08, hunt10}; see also the discussions of PAHs from \citealt{cormier15,cormier19} and the references therein).  This was interpreted as evidence that PAH emission was dependent on metallicity.  However, the relative strength of PAH emission appears to depend more closely on the  hardness or intensity of the illuminating radiation field \citep{engelbracht08, gordon08, hunt10}.  Since low metallicity environments contain less dust, the interstellar radiation field should be harder and more intense in these regions, which are conditions that may lead to suppressed PAH emission.  Similarly, since molecular gas is expected to be less prevalent in low metallicity regions or regions with harder radiation fields, the molecular gas will provide less shielding to the PAHs in these locations.   Related to these findings, \citep{bendo08} found that the gradient in the PAH/160~$\mu$m emission ratio does not correlate with the gradient in log(O/H), which implied that PAH excitation was not primarily linked to metallicity.

Ideally, it would be good to follow up this analysis with a comparison of PAH/dust ratios to metallicity measured within individual subregions within individual galaxies, especially since abundances may vary azimuthally as well as radially \citep[e.g. ][]{sanchezmenguiano16, ho17, lin17, kreckel19}.  At this point in time, however, such images are exceptionally difficult to create and are generally not publicly available.  We can at least re-examine the relation between the PAH/dust gradients and metallicity gradients using our data.  If metallicity variations are a strong driver of variations in PAH emission, the two gradients should appear related.

\begin{figure}
\epsfig{file=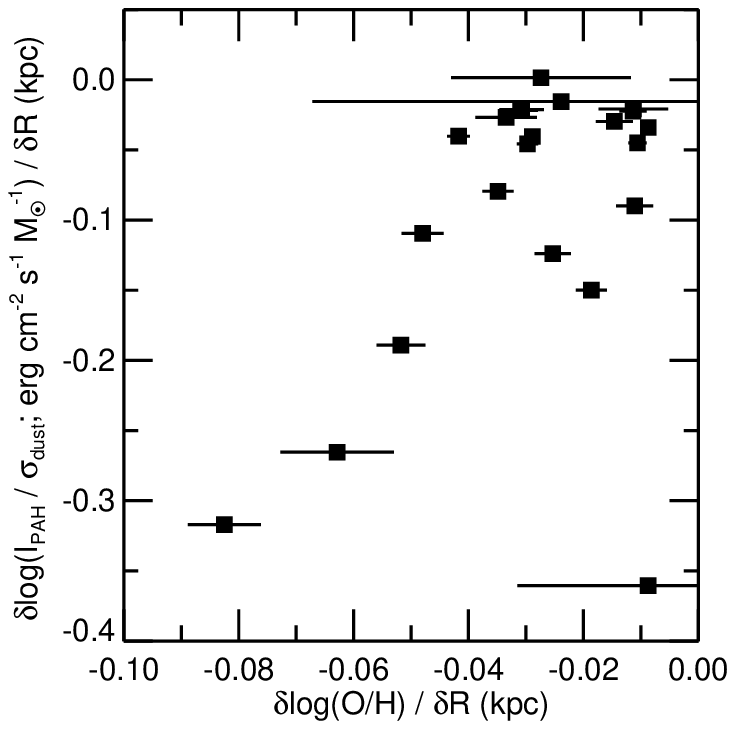}
\caption{Plot of the gradient of the logarithm of the PAH/dust ratio versus the gradient in log(O/H).  The data used in these plots are listed in Table~\ref{t_grad}.}
\label{f_grad}
\end{figure}

Table~\ref{t_grad} lists the galactocentric radial gradients in the logarithm of the PAH/dust ratios (as measured from the data with 8~arcsec pixels and matching PSFs but no additional smoothing) to the gradients measured in log(O/H) by \citet{pilyugin14}, who provide homogeneous data for 22 of the galaxies in our sample.  The gradients are plotted versus each other in Figure~\ref{f_grad}.  For this comparison, we measured gradients using the same disc position angles and inclinations given by \citet{pilyugin14}, although most values for the position angles and all inclination angles vary by less that 10~degrees.  However, we rescale the gradients in log(O/H) to correspond to the distances listed in Table~\ref{t_dist}.

NGC~300, NGC~2403, and NGC~7793, which have the largest log(O/H) gradients, also exhibit large PAH/dust gradients, which indicates that metallicity could play a role in the PAH excitation in these three galaxies.  However, NGC~4736 has the steepest PAH/dust gradient but a relatively shallow log(O/H) gradient, which indicates that the enhancement of PAH emission in the inner region of this galaxy is much more likely to be related to the brighter stellar populations than to any metallicity variations.  All other galaxies plotted in Figure~\ref{f_grad} show no clear trend.  The weighted correlation coefficient for all of the data is 0.55, which would indicate that $\sim$30\% of the variance in the PAH/dust gradient can be described as a function of the log(O/H) gradient.  If we exclude NGC~4736, the weighted coefficient increases to 0.62.  If we exclude the four galaxies with the steepest log(PAH/dust) gradients, the weighted correlation coefficient becomes 0.17.

The tests in Appendix~\ref{a_disttest} indicate that distance-related blurring effects could cause the gradient in the logarithm of the PAH/dust ratio to vary by $\sim$20\% with a 3$\times$ variation in distance.  However, if we select subsets of the data by distance (such as galaxies between 5 and 15~Mpc or 10 and 20~Mpc), our interpretation does not change.  Having said this, the galaxies with the steepest PAH/dust gradients are all within 5~Mpc, which may be indicative of a bias in our sample selection criteria against including galaxies with such steep gradients that are at larger distances.  Additional data could reveal NGC~4736 to be an outlier in the relation between PAH/dust and log(O/H) gradients or could show that the trend seemingly shown by NGC~300, NGC~2403, and NGC~7793 in Figure~\ref{f_grad} does not apply to other galaxies with steep PAH/dust gradients.

None the less, we can state that the gradients of the PAH/dust ratio could correlate with metallicity (or at least the gas-phase metallicity traced by log(O/H)) in situations where the metallicity gradients are relatively strong.  However, this does not necessarily prove any causal link between the two gradients, and more detailed analyses of data, preferably within subregions of these and similar galaxies, would be needed to examine the relations further.  In most other galaxies, the observed variations in the PAH/dust ratio appear to be independent of the metallicity gradient, implying that metallicity has no direct effect on PAHs in most galaxies.
 
\begin{table}
\caption{Radial gradients in log($I_{PAH})/\sigma_{dust}$) and log(O/H)}
\label{t_grad}
\begin{center}
\begin{tabular}{@{}lcc@{}}
\hline
Galaxy &
    Gradient in &
    Gradient in \\
&
    log(O/H)$^a$ &
    log($I_{PAH}/\sigma_{dust}$)$^b$\\
&
    (dex kpc$^{-1}$) &
    (log(erg cm$^{-2}$ s$^{-1}$ M$_\odot^{-1}$) kpc$^{-1}$) \\
\hline
NGC 300  &
    -0.082$\pm$0.006 &
    -0.317$\pm$0.002 \\
NGC 628  &
    -0.042$\pm$0.002 &
    -0.040$\pm$0.001 \\
NGC 925  &
    -0.035$\pm$0.003 &
    -0.079$\pm$0.001 \\
NGC 1097 &
    -0.011$\pm$0.002 &
    -0.045$\pm$0.001 \\
NGC 1365 &
    -0.009$\pm$0.001 &
    -0.034$\pm$0.001 \\
NGC 2403 &
    -0.052$\pm$0.004 &
    -0.189$\pm$0.001 \\
NGC 3031 &
    -0.011$\pm$0.003 &
    -0.090$\pm$0.001 \\
NGC 3184 &
    -0.011$\pm$0.002 &
    -0.022$\pm$0.001 \\
NGC 3351 &
    -0.019$\pm$0.003 &
    -0.150$\pm$0.001 \\
NGC 3621 &
    -0.048$\pm$0.004 &
    -0.110$\pm$0.001 \\
NGC 3938 &
    -0.033$\pm$0.005 &
    -0.027$\pm$0.001 \\
NGC 4254 &
    -0.031$\pm$0.003 &
    -0.022$\pm$0.001 \\
NGC 4303 &
    -0.031$\pm$0.004 &
    -0.021$\pm$0.001 \\
NGC 4321 &
    -0.015$\pm$0.003 &
    -0.030$\pm$0.001 \\
NGC 4501 &
    -0.027$\pm$0.016 &
     0.001$\pm$0.001 \\
NGC 4725 &
    -0.024$\pm$0.043 &
    -0.016$\pm$0.001 \\
NGC 4736 &
    -0.009$\pm$0.023 &
    -0.361$\pm$0.001 \\
NGC 5055 &
    -0.030$\pm$0.002 &
    -0.046$\pm$0.001 \\
NGC 5236 &
    -0.025$\pm$0.003 &
    -0.124$\pm$0.001 \\
NGC 5248 &
    -0.011$\pm$0.006 &
    -0.021$\pm$0.001 \\
NGC 5457 &
    -0.029$\pm$0.001 &
    -0.040$\pm$0.001 \\
NGC 7793 &
    -0.063$\pm$0.010 &
    -0.265$\pm$0.001 \\
\hline
\end{tabular}
\end{center}
$^a$ These gradients are from \citet{pilyugin14}, but they have been rescaled to correspond to the distances listed in Table~\ref{t_dist}.\\
$^b$ The gradients were measured using the definitions of the position angles and inclinations given by \citet{pilyugin14} but using the centres of the galaxies listed in Table~\ref{t_sample} and the distances listed in Table~\ref{t_dist}.\\
\end{table}

\section{Discussion}
\label{s_discuss}

To summarize the analysis results, PAH excitation as traced by the PAH/dust ratios appears linked to different stellar populations in different galaxies.  In 11 of the 25 galaxies we examined, the PAH excitation appeared straightforwardly linked to star formation activity and was more often more strongly linked specifically to 24~$\mu$m hot dust emission tracing the energy from star forming regions that is absorbed by dust.  In another 5 of the sample galaxies, the PAH emission appeared correlated with star formation activity only parts of the galaxies' discs.  In another 6 galaxies, the PAH excitation was more strongly linked to 3.6~$\mu$m emission that originates primarily from evolved stars.  The final 3 galaxies produced ambiguous results.  This myriad of results shows that PAH excitation cannot be described as linked to a single excitation source in all galaxies.

\subsection{A description of PAH excitation around star forming regions}

Although, the PAH/dust ratio is strongly correlated to one or more star formation tracers in parts of most galaxies (at the $\sim$1~kpc resolutions of the data in our analysis), PAH emission itself still appears low compared to other star formation tracers within the centres of star forming regions and also appears high compared to star formation tracers in diffuse regions \citep{calzetti05, prescott07, bendo08}.  To reconcile these two phenomena, we suggest that the PAHs tend to be associated with warm dust (dust cooler than the dust at the centres of these regions but hotter than dust in the diffuse ISM) and molecular gas within shells around these star forming regions.  This would be consistent with observations of PAHs in shells around star forming regions within the Milky Way \citep{povich07, kirsanova08, watson08, dewangan13} and in other galaxies \citep{helou04, bendo06}.  The radiation fields within the centres of the star forming regions are potentially strong and hard enough to destroy PAHs.  However, dust and molecular gas within the inner layers of the dust shells may shield the PAHs from the hardest ultraviolet photons, while softer ultraviolet and optical photons can potentially penetrate further into these shells and excite the PAHs.  This leads to enhanced PAH emission relative to the cold dust emission in the shells that appears correlated to star formation when averaged over kiloparsec-sized areas.  This scenario would also explain how PAH emission itself does not correlate well with hot dust emission at 24~$\mu$m but does correlate with emission at longer infrared wavelengths that would be expected to originate from dust within these shells \citep{bendo08, verley09, jones15, cortzen19}.
 
When the PAH/dust ratios are associated with star formation, they sometimes correlate equally with both far-ultraviolet and mid-infrared emission.  However, the ratios correlate significantly better with mid-infrared emission in several galaxies, suggesting that the PAHs in these galaxies are excited by ultraviolet photons that ultimately do not escape the outer dust shells.  Conversely, the ratios correlate better with far-ultraviolet emission that mid-infrared emission in the inner regions of NGC~5236 and NGC~5457.  These could be cases where the PAHs are found in locations with relatively less dust.  The ultraviolet emission may not be heavily obscured, but at the same time, the dust attenuation is not so low that the PAHs are exposed to very hard ultraviolet photons.  Alternately, \citet{jones15} suggested scenarios for NGC~5236 where the PAHs could be excited by ultraviolet light from a young but non-photoionizing population or by ultraviolet light diffusing from the spiral arms, but it is not clear that these scenarios would apply to NGC~5457.  More detailed radiative transfer modelling may be needed to understand these variations in how PAH excitation is associated with different star formation tracers in different circumstances. 

\begin{figure}
\epsfig{file=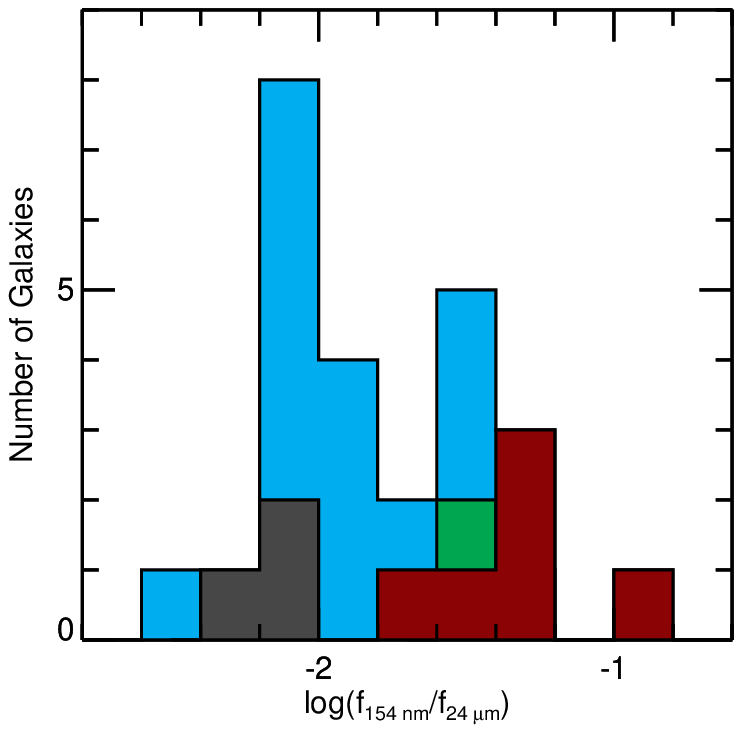}
\caption{A histogram of the logarithm of the $f_{154nm}/f_{24\mu m}$ ratio for the sample galaxies. The subset of galaxies where the PAH excitation is most strongly linked to the evolved stellar population (or where the PAH/dust ratios are more strongly correlated with 3.6~$\mu$m emission) are coloured red.  
The subset of galaxies where the PAH excitation is most strongly linked to the evolved stellar population (or where the PAH/dust ratios are more strongly correlated with one or more star formation tracers) are coloured cyan.  Galaxies that produced ambiguous results are coloured grey.  NGC~5457, where PAH excitation was linked to the evolved stellar population in the outer part of the galaxy disc, is coloured green.} 
\label{f_fuv24histogram}
\end{figure}

\subsection{Circumstances where PAH excitation may be linked to the evolved stellar population}
\label{s_discuss_pahevolvedstars}

Five of the six galaxies where the PAH excitation is more strongly linked to the evolved stellar populations (or where the PAH/dust ratios are more strongly correlated with 3.6~$\mu$m emission) all appear very distinctly different from the rest of the sample in terms of their stellar and infrared morphologies.  NGC~300, NGC~925, NGC~2403, NGC~3621, and NGC~7793 are all classified as Scd or Sd galaxies, which means that they have negligible bulges ad loosely-wound spiral arms.  M33, where the PAH excitation was linked to evolved stars by \citet{calapa14}, is also an Scd galaxy \citep{devaucouleurs91}, and NGC~5457, where the PAH excitation appeared linked to evolved stellar populations in the outer disc, is also an Scd galaxy.  These results would imply that some characteristics of late-type spiral galaxies results in PAHs being excited by evolved stars.  However, NGC~3184 and NGC~4303, where PAH excitation is linked to star formation, are also Scd galaxies, and PAH excitation is also linked to evolved stars in NGC~3031, which is an Sab galaxy. 

We also found that, regardless of the morphological classification of the galaxy, the ratio of globally-integrated far-ultraviolet to mid-infrared emission (or $f_{154 nm}/f_{24 \mu m}$) is much higher in galaxies where the PAH excitation is linked to the evolved stellar populations.  Figure~\ref{f_fuv24histogram} shows a histogram of the logarithm of the globally-integrated $f_{154 nm}/f_{24 \mu m}$ ratios for our sample galaxies.  

The galaxies where the PAH/dust ratios are more strongly correlated to 3.6~$\mu$m emission than to any star formation metric, which are coloured red, lie at the right side of this histogram.  The galaxies where the PAH/dust ratios are more strongly correlated to star formation, which are coloured blue, lie in the centre and to the left.  A separate green box for NGC~5457, where the PAH/dust ratios were linked to  star forming regions in the inner disc but the evolved stellar population in the outer disc, also lies on the right side of the distribution of $f_{154 nm}/f_{24 \mu m}$ values.  NGC~5457 might in fact be transitory between the galaxies in cyan and the galaxies in red, which may explain its placement on the graph.  Interestingly, the change from PAH excitation linked to star formation in the inner disc of NGC~5457, where the 154~nm/24~$\mu$m ratio is low, to PAH excitation linked to evolved stars in the outer disc, where the 154~nm/24~$\mu$m ratio is high, is consistent with the overall difference seen between the red and blue distributions in Figure~\ref{f_fuv24histogram}.  The three galaxies where the results are ambiguous, which are shown in grey, lie within the distribution of the cyan data points.  While the infrared emission is high relative to the ultraviolet emission in both sets of galaxies, it would not be safe to extrapolate from this that, in the galaxies with ambiguous results from our analysis, the PAHs are actually excited by star forming regions.

Applying a Kolmogorov-Smirnov test, we find that the likelihood of the galaxies shown in blue and red in Figure~\ref{f_fuv24histogram} coming from different populations is 99.5\%.  If NGC~5457 is treated as part of the blue subset, that percentage changes to 98.2\%, but the percentage becomes and 99.8\% when NGC~5457 is treated as part of the red distribution.

These results provide clear empirical evidence that PAH excitation appears linked to evolved stellar populations when $f_{154 nm}/f_{24 \mu m}$ is high.  This suggests that dust extinction is related to how PAHs are excited, but the mechanism linking these two phenomena is unclear.  One possible explanation is that PAH excitation is normally constrained to locations near star forming regions in galaxies with average to high dust extinction levels because the interstellar radiation field is very red except near star forming regions, but in galaxies with low dust extinction, the interstellar radiation field will appear bluer overall.  Hence, in these low dust extinction galaxies, the evolved stars could contribute more to exciting PAHs, thus making the PAH/dust ratio correlate with 3.6~$\mu$m emission.  Unfortunately, this does not quite explain why the PAH/dust ratios should appear decoupled from star formation tracers in galaxies with high $f_{154 nm}/f_{24 \mu m}$ ratios.  Another possibility is that, in galaxies with low dust densities, the photons that excite PAHs are not absorbed close to the star forming regions.  Hence, the PAHs will not appear locally excited around star forming regions but could appear excited areas extending beyond the clouds surrounding these star forming regions into the diffuse ISM.  However, it is not entirely clear that such a diffusion phenomenon could make the PAH/dust ratios appear more closely linked to 3.6~$\mu$m emission.  If this diffusion phenomenon is coupled with more excitation of PAHs by evolved stars, it could explain how the PAH/dust ratios are more strongly correlated to 3.6~$\mu$m emission in galaxies with high $f_{154 nm}/f_{24 \mu m}$ ratios.  Additional radiative transfer modelling would be able to verify this scenario for PAH excitation within galaxies.  Observations of molecular gas, which also play a role in shielding PAHs, could potentially illustrate its role in altering how PAHs are excited, but such an analysis is beyond the scope of this paper.

\section{Conclusions}
\label{s_conclu}

To summarize, we have used the ratio of mid-infrared PAH emission to dust surface density to study how PAHs are excited by star forming regions and the evolved stellar populations in 25 nearby face-on spiral galaxies.  We find that the nature of PAH excitation varies among galaxies, with different stellar populations appearing to be responsible for exciting PAHs in different galaxies.

In 11 of the 25 galaxies, we found that PAH excitation was linked to star formation across the entire disc of each galaxy. In another 5 galaxies, PAH excitation was enhanced around star forming regions only within specific ranges of galactocentric radii.  Prior results show that PAH emission appear suppressed relative to other star formation tracers within star forming regions \citep{calzetti05, prescott07, bendo08} and that PAHs are typically seen in shell-like structures around star forming regions \citep{helou04, bendo06, povich07, kirsanova08, watson08, dewangan13}.  Our results along with these other observations suggest that the PAH are most strongly excited in dusty shells around star forming regions that can be penetrated by soft ultraviolet photons from the star forming regions but that are shielded from harder photons that either destroy the PAHs or otherwise inhibit the PAH emission.

In another 6 galaxies, PAH excitation was linked to the evolved stellar population throughout the entire disc of each galaxy, and in NGC~5457, PAH excitation was linked to evolved stars in the outer disc.  The high 154~nm/24~$\mu$m ratios for these galaxies indicates that dust extinction is lower in these galaxies.  We suggest that lower dust extinction around star forming regions will cause the PAH/dust ratios to appear decoupled from star formation tracers.  At the same time, diffuse ultraviolet light, including ultraviolet photons escaping from these star forming regions and ultraviolet photons within the interstellar radiation field from evolved stars, can propagate further throughout the discs of these galaxies and can therefore dominate PAH excitation.  

The remaining 3 galaxies produced ambiguous results.  This is a consequence of the morphologies of the galaxies looking similar in 3.6~$\mu$m emission and at least one of the star formation tracers.

The overall results from the sample of 25 galaxies have complex implications for using PAH emission as a global star formation tracer.  Since PAHs are excited by star forming regions in most but not all spiral galaxies, PAH emission functions as a typical star formation tracer in the sense that it traces the energy output of the star forming regions.  However, galaxies in which PAHs are excited by the evolved stellar population would be expected to add scatter to the relation between globally-integrated PAH emission and star formation.  This would be similar to how \citet{calzetti10} found that dust continuum emission exhibited more scatter relative to other star formation tracers when observed at longer wavelengths because of the transition from dust heated primarily by star forming regions to dust heated primarily by evolved stars.  Having said this, \citet{boquien11} and \citet{bendo12a} demonstrated that dust heated primarily by evolved stars is correlated to other star formation metrics because the dust emission is related to gas mass and gas mass correlates with star formation rate \citep{schmidt59, kennicutt98}; the same principle could apply when PAHs are primarily excited by evolved stars.  In any case, the relation between PAH emission and star formation is clearly not as straightforward as it is for other bands.

The overall results regarding PAH excitation also have implications for dust emission and radiative transfer modelling.  Multiple models have incorporated PAHs (or other sources of the mid-infrared spectral features) into their models \citep{silva98, draine07a, bianchi08, baes11, jones13, domingueztenreiro14}, and some models have even been used to predict whether the PAH emission originates from areas near star forming regions or the diffuse ISM \citep{draine07b, popescu11}.  The models need to be tuned to account for how PAHs frequently appear to be excited in locations near but not within star forming regions.

\section*{Acknowledgments}

We thank the reviewer for their comments on this paper.  GJB acknowledges support from STFC Grant ST/P000827/1 and STF Grant ST/T001488/1. This research has made use of the NASA/IPAC Extragalactic Database (NED), which is operated by the Jet Propulsion Laboratory, California Institute of Technology, under contract with the National Aeronautics and Space Administration. 

{}

\appendix

\section{Tests of dust mass calculations based on long-wavelength far-infrared data}
\label{a_disttest}

\begin{table*}
\centering
\begin{minipage}{176mm}
\caption{Statistics for relation of log($I_{PAH}/\sigma_{dust}$) to other quantities when using different bin sizes}
\label{t_disttest}
\begin{tabular}{@{}lccccccc@{}}
\hline
Galaxy &
  Bin Size &
  Number &
  \multicolumn{4}{c}{Correlation Coefficient for Relations of Quantities to log($I_{PAH}/\sigma_{dust}$)} &
  Slope in log($I_{PAH}/\sigma_{dust}$) \\

&
  (arcsec) &
  of Bins &
  log($I_{154 nm}$) &
  log($I_{154nm + 24\mu m}$) &
  log($I_{24 \mu m}$) &
  log($I_{3.6 \mu m}$) &
  versus radius (kpc$^{-1}$) \\
\hline
NGC 3031 &           
    24 &    652 &    -0.11 &     0.41 &     0.68 &     0.92 &
    $-0.090\pm0.001$ \\
    &
    72 &     81 &    -0.14 &     0.47 &     0.77 &     0.94 &
    $-0.086\pm0.002$ \\
NGC 5457 &           
    24 &    657 &     0.46 &     0.67 &     0.78 &     0.88 &
    $-0.040\pm0.001$ \\
    &
    72 &    111 &     0.63 &     0.79 &     0.87 &     0.92 &
    $-0.050\pm0.001$ \\
NGC 5457 ($r$$\leq$5~kpc) &           
    24 &    112 &     0.86 &     0.85 &     0.75 &     0.46 &
    \\
    &
    72 &     11 &     0.90 &     0.93 &     0.86 &     0.41 &
    \\
NGC 5457 ($r$$>$5~kpc) &           
    24 &    545 &     0.44 &     0.60 &     0.73 &     0.92 &
    \\

\hline
\end{tabular}
\end{minipage}
\end{table*}

As seen in Table~\ref{t_dist}, the distances to the galaxies in our sample as well as the physical area covered by the 24~arcsec bins that we use for sampling regions within these galaxies varies by a factor of 10.  Star forming and diffuse regions will naturally appear more blended for more distant galaxies, which could affect the results.

As a test of the robustness of our analysis against distance effects, we simulated what would happen with NGC~3031 and NGC~5457 (the two galaxies with the largest angular sizes in our sample and the galaxies where we had the most 24~arcsec bins in our analysis) if we used larger bin sizes to sample the data.  We found that the galaxies would still have enough bins to meet our sample selection criteria if we increased the bin size 3$\times$ (to 72~arcsec) but not larger.  This would be equivalent to moving NGC~3031 from $\sim$3.6 to $\sim$10.8~Mpc, which would shift it from the short end to the midrange of distances for our sample, and it would be equivalent to moving NGC~5457 from 6.95~Mpc to 20.85~Mpc, which would be equivalent to shifting the galaxy to the high end of the distances in our sample.  The smoothing also leads to detections at the 3$\sigma$ level over a larger area of each galaxy, but we did not make any adjustments for this effect.

Table~\ref{t_disttest} provides statistics on the correlation coefficients using the 24 and 72~arcsec bins.  All of the correlation coefficients increase in value when the bin size is increased, which would be expected given that the larger bin size smooths the data.  However, the relative differences between the correlation coefficients do not change.  In NGC~3031, we would still conclude that the PAH/dust ratios are most strongly correlated with 3.6~$\mu$m emission.  In NGC~5457, we would still conclude that the ratios are more strongly correlated with any star formation tracer than with 3.6~$\mu$m emission at $r$$\leq$5~kpc but that the ratios are most strongly correlated with 3.6~$\mu$m emission at $r$$>$5~kpc (although we only have 11 data points in the coarsely-binned data at $r$$\leq$5~kpc, so those results would be treated with caution in our analysis).  This illustrates that the parts of our analysis relying on correlation coefficients should be robust to some degree against distance-related effects as long as we do not compare coefficients for relations measured for different galaxies.  However, binning the data further would cause more difficulty in identifying differences in the relations between PAH/dust ratios and the tracers of different stellar populations within individual galaxies.

We also examined how the radial profiles of the logarithms of the PAH/dust ratios changed from the measurements in the maps with 8~arcsec pixels and 25~arcsec PSFs compared to the 72~arcsec binned data.  These numbers, which are also listed in Table~\ref{t_disttest}, show that the smoothing related to a 3$\times$ change in distance could alter the measured gradient by 20\%.  Although this change was only seen in NGC~5457, this could be a general concern when comparing the gradients of the PAH/dust ratio to log(O/H) gradients in Section~\ref{s_radialvar}.

\section{Tests of dust mass calculations based on long-wavelength far-infrared data}
\label{a_dustmass}

As we discussed in Section~\ref{s_data}, we needed a method of calculating dust mass that was relatively simple but that could provide reasonably accurate values.  Although far-infrared emission from a subregion within a galaxy may originate from dust with a broad range of temperatures, \citet{bianchi13} has argued that using a single modified blackbody fit to $>$100~$\mu$m data could produce reasonably accurate dust mass measurements.  \citet{peretto16} have also shown that using just the 160 and 250~$\mu$m bands by themselves is a viable way to calculate dust masses, at least for infrared dark clouds.  However, since \citet{bendo12a} and \citet{bendo15} showed that extragalactic emission at 160 and 250~$\mu$m is likely to originate from different thermal component of dust heated by different stellar populations, we would prefer to use data at wavelengths longer than 160~$\mu$m.  We found through experimentation that using a modified blackbody fitted to just 250 and 350~$\mu$m data points could reproduce dust masses over a broad range of scenarios.  

To test how this works, we created a series of four different types of models more complex than a single modified blackbody.  Each of these models have been used frequently to describe dust emission in the far-infrared and to measure dust masses.  The first two models are based on the addition of two modified blackbodies; the first has a mass ratios of 1/10, and the second has a mass ratio of 1/2.  The warmer component represents dust heated locally by star formation, while the colder component represents cirrus dust heated by diffuse starlight.  The third model is based on the sum of a range of modified blackbodies in which the mass of a given modified blackbody in the sum scales as $U^{-\alpha}$, where $U$ is the intensity integrated over wavelength of the radiation field illuminating the thermal component (or the integral over wavelength of the emission from the dust) and $\alpha$ is an index typically fit to the data.  This describes dust located in an assortment of clouds or other structures with intensities that scale as a power law and is used, for example, in the SED templates created by \citet{dale02}.   The final model is a variant of the third model in which the coldest modified blackbody is treated as a separate cirrus component and is scaled to a larger value than given by the power law relation.  For this fourth model, the ratio of the mass in the sum of the warmer thermal components to the mass of the single cold thermal component is set to 1/10.

When building these SEDs, we used several constraints on the data.   In our 24~arcsec binned data, we only observed 70/100, 70/160, ad 100/160~$\mu$m colour temperatures between 15 and 36~K, and we only observed 160/250 and 250/350~$\mu$m colour temperatures between 10 and 36~K.  We also observed differences of $<$12~K between the 70/100 and 250/350~$\mu$m colour temperatures or between the 100/160 and 250/350~$\mu$m colour temperatures.  We therefore only analyzed models with colour temperatures that matched these ranges.

In the two temperature component scenarios, we required that the colder dust must produce at least 20\% of the expected flux at 500~$\mu$m.  If the warmer dust component is too bright compared to the colder dust component, (for example, when the coldest dust is $\sim$10~K and the warmer dust is $>$25~K), the colder dust component would be very difficult to detect with {\it Herschel} or any other infrared telescope.  In such situations, any type of SED fitting or radiative transfer modelling will have difficulty accounting for the presence of the cold dust component, as it represents such a small fraction of the energy absorbed and emitted by dust.  Moreover, random fluctuations in data where the signal-to-noise ratios are $\sim$5 could potentially mimic the presence of a very cold but very massive dust component that contributes $\sim$20\% of the emission in the longest infrared bands.

When we scaled the thermal components by a power law, we used $\alpha$ values that ranged between 1.5 and 4.0, which was typical for the range of $\alpha$ values for the models fit to nearby spiral galaxy SEDs by \citet{dale07}.  The range of $U$ values varied by a factor of 20000, which is similar to what is used by \citet{draine07b} in their physical dust models. 

\begin{figure}
\begin{center}
\epsfig{file=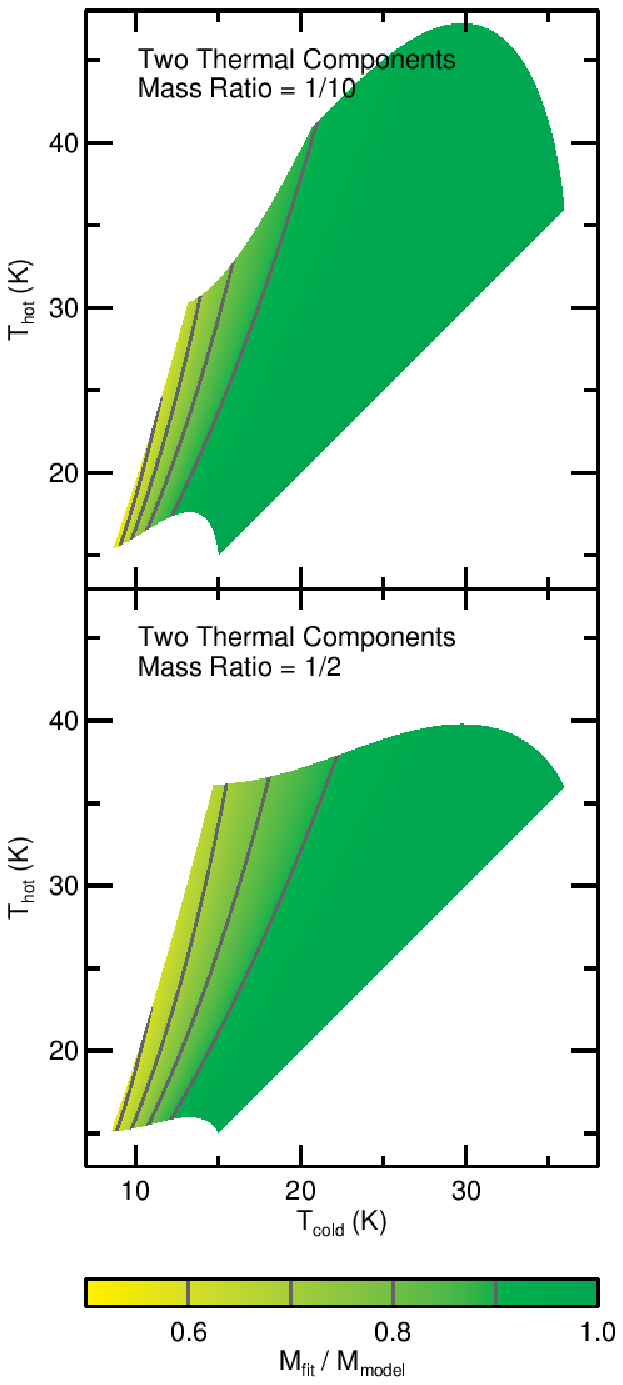}
\end{center}
\caption{Plots of the ratio of the dust mass measured from a single modified blackbody fit to the 250 and 350~$\mu$m data versus the actual dust mass for two models based on the sums of two modified blackbodies.  White areas in these figures are locations where the models had colours that were inconsistent with our data, where $T_{cold}$$>$$T_{hot}$, or where $<$20\% of the emission at 500~$\mu$m would originate from the cold component, thus making it virtually impossible to detect.}
\label{f_masstest_mult2tmodel}
\end{figure}

\begin{figure}
\begin{center}
\epsfig{file=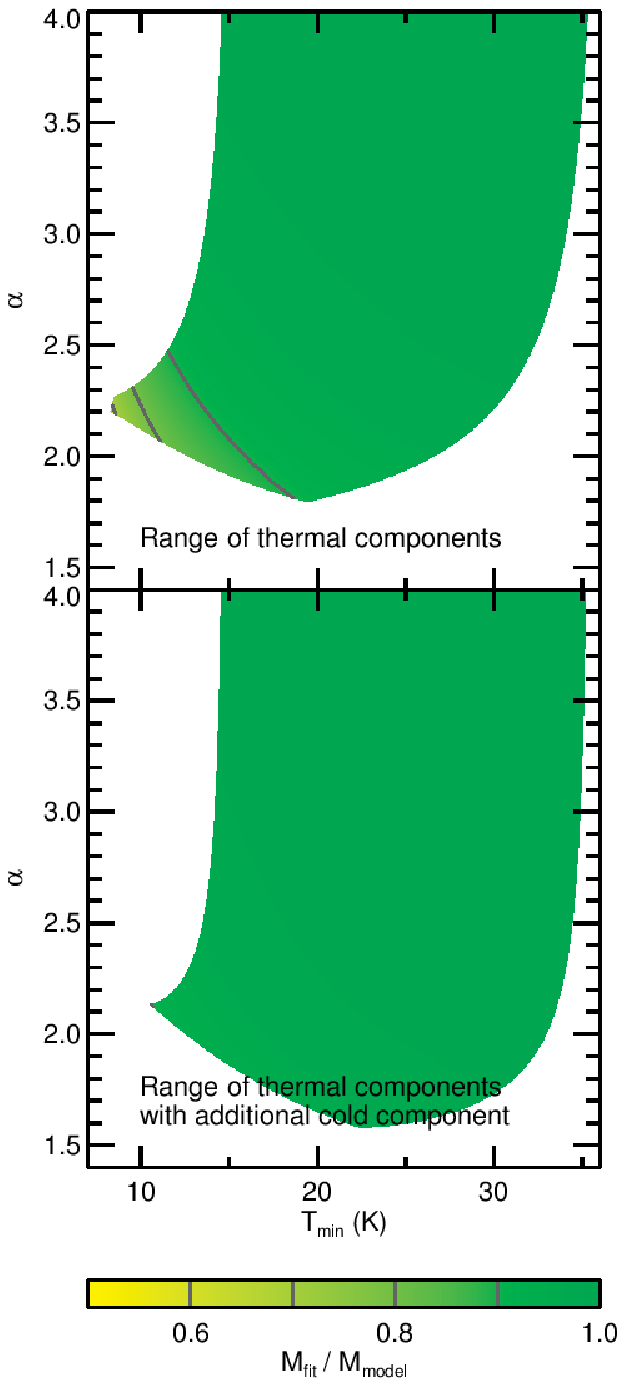}
\end{center}
\caption{Plots of the ratio of the dust mass measured from a single modified blackbody fit to the 250 and 350~$\mu$m data versus the actual dust mass for two complex sums of modified blackbodies.  The top panel is for a model with a range of thermal components where the masses of the component scale as $U^{-\alpha}$.  The bottom panel show the results for a similar power law model where the coldest thermal component is scaled independently of the other components.  White areas in these figures are locations where the models had colours that were inconsistent with observations of nearby spiral galaxies.}
\label{f_masstest_multpowerlawmodel}
\end{figure}

We creates SEDs based on each of these models using modified blackbodies with dust emissivities based on the values from \citet{draine03}, which are approximately proportional to $\nu^2$.  We then fit the 250 and 350~$\mu$m data points with a single modified blackbody.  (While in Section~\ref{s_data_addcalc} we noted that it was sometimes necessary to use the shorter wavelength data because the function fit to the 250 and 350~$\mu$m data superseded the measurements in those bands, the modified blackbody fit to the 250 and 350~$\mu$m data from these theoretical functions always fell below the measurements at shorter bands.)  We then compared the mass of the fitted modified blackbody to the mass of the dust in the more complex input SED.  The resulting ratios of the fitted masses to the input masses are shown in graphical form in Figures~\ref{f_masstest_mult2tmodel} and \ref{f_masstest_multpowerlawmodel}.

In the scenarios with two modified blackbodies, the mass from the 250 and 350~$\mu$m data is within 20\% of the actual mass in most situations where the temperatures of the two components differ by $<$5~K or when the mass ratio is 1/10 and the temperature of the colder component is $\gtsim$15~K.  Values within 20\% are obtained over most of the explored range, which leads to a variation of $<$0.10 in logarithm space.  This would be acceptable for dust surface density measurements in our analysis given that log($I_{PAH}$/$\sigma_{dust}$) varies by at least 0.5 and often more than 1.0 for the galaxies in our analysis.  It would also be significantly better than using a single band as a proxy for dust mass.  For example, the 250~$\mu$m flux density could vary by $\sim$8$\times$ for dust temperature variations between 15 and 30~K.  

The 250 and 350~$\mu$m data give dust masses that differ by more than 20 \% from the two temperature component models when the dust temperatures differ by $>$5~K and the temperature of the colder dust is $\ltsim$15~K.  The 24~arcsec bins in our sample galaxies that would be best described by such two temperature components are located at the periphery of the detected infrared emission where the signal-to-noise ratios are low and where we may not have $>$3$\sigma$ detections in the 70 or 100~$\mu$m data.  In these situations, random fluctuations in the longer wavelength data sometimes give the illusory appearance that a large mass of cold dust is present, although random fluctuations can also make the data appear more consistent with a single modified blackbody.  Our reliance on weighted correlation coefficients should help us to avoid the uncertainties in estimating the dust masses in these regions.  Having said that, dust mass estimates of 60\% of the total mass (which is near the limit of what is seen for any scenario in Figure~\ref{f_masstest_mult2tmodel}) would cause a change of $\sim$0.22 in the logarithm of the PAH/dust ratios.  This is less than the dynamic range of the observed values in any of our sample galaxies, but scatter on this scale would make it more difficult to discern differences between the relations of PAH/dust ratios to either star formation tracers or 3.6~$\mu$m emission.  This could affect our analysis in low surface brightness regions where the colours match these type of extreme two temperature component models, but most of our data should not be as strongly affected.

As an additional test of the two modified blackbodies, we constructed models based on the results from \citet{bendo15} where the SEDs for three galaxies (NGC~628, NGC~2403, and NGC~5457) were separated into separate thermal components using the colour variations within the galaxies  (although the mass and temperature of the warmer component was not treated as physically meaningful in that analysis, as dust grains with a range of temperatures probably produce the emission described by the single warm component).  When we fit just the 250 and 350~$\mu$m data points in the model SEDs, the estimated masses are within 15\% of those from the two separate components.

The 250 and 350~$\mu$m data generally produced masses that were within 10\% of the actual model masses in most cases where the SEDs were constructed using a power law relation.  The colour temperatures from the 250/350~$\mu$m ratios are usually very similar to the coldest dust temperatures, which leads to more accurate measurements of the mass.  The 250 and 350~$\mu$m data did not yield masses that were within 20\% of the model masses only for scenarios with no additional cold component (the top panel of Figure~\ref{f_masstest_multpowerlawmodel}) where the coldest temperatures are $\ltsim$10~K and where the $\alpha$ is $\sim$2.2.  These are scenarios where the colour temperatures differ significantly between 70 and 350~$\mu$m and are near the extremes in colour temperature variations that we observe in our sample.  The 24~arcsec binned data that are best fit by these extreme SEDs are again low signal-to-noise locations near the edges of the detected regions in our sample galaxies where 1$\sigma$ fluctuations can lead to large variations in the shape of the SED.  Even so, the most discrepant models still produce mass measurements within 30\% of the actual model mass.  This equivalent to a change of 0.15 in logarithm of the PAH/dust ratio, which is well within the dynamic range of the data for any galaxy and should be less of a concern than the issues with the two thermal component models.

\begin{figure*}
\epsfig{file=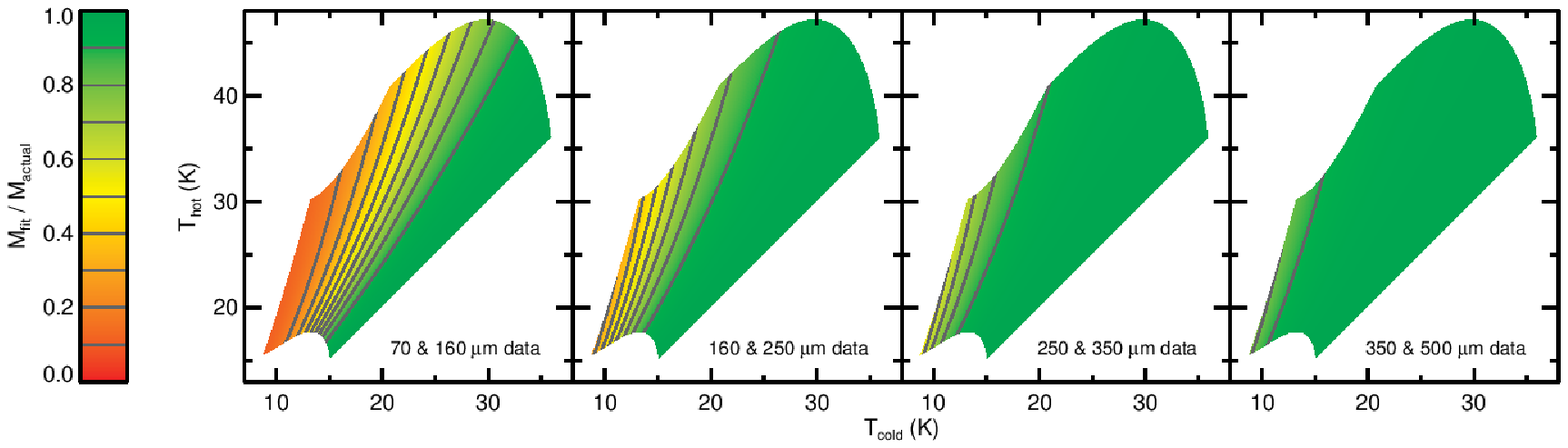}
\caption{Plots of the ratio of the dust mass measured from a single modified blackbody fit to two wavebands versus the actual dust mass for a model based on the sum of two modified blackbodies with a mass ratio of 1/10. Areas are displayed in white for the same reasons as in Figure~\ref{f_masstest_mult2tmodel}.}
\label{f_masstest_multtc}
\end{figure*}

The one scenario where we cannot accurately estimate dust masses using just the 250 and 350~$\mu$m data is the situation in which dust emissivity index $\beta$ is variable.  If this happens, then calculating accurate dust masses in general becomes very difficult.  The first problem is that the emissivity is assumed to be fixed to a value from a physical model, such as the values from \citet{draine03}, at a specific wavelength, such as 250~$\mu$m \citep[e.g. ][]{galametz12, hunt15} or 350~$\mu$m \citep[e.g. ][]{smith12}, but this choice is ultimately arbitrary, and it can affect the resulting dust masses.  The second problem is that, when $\beta$ is allowed to vary in SED fits, an inverse correlation between $\beta$ and temperature can appear because of noise or because of the mixing of multiple thermal components along the line of sight \citep{shetty09, galametz12}.  Some analyses on nearby galaxies suggest that variations in $\beta$ are caused by mixing thermal emission from dust heated locally by star forming regions and diffuse dust heated by evolved stars \citep{kirkpatrick14, hunt15}, and adjusting the emissivity function could significantly change the resulting dust temperature and mass and produce inaccurate measurements of these quantities.  \citet{galametz12} also found that, when using a variable $\beta$ to fit subregions within nearby galaxies, temperature spikes would appear in regions with no obvious dust heating sources, which indicates that the results from these fits may not be reliable.  Given these results, we think it is very unlikely that $\beta$ is actually variable and that fits based on a variable $\beta$ do not yield reasonably accurate dust masses.  

As a brief evaluation of how the choice of {\it Herschel} or {\it Spitzer} wavebands influences the results, we calculated dust masses for two modified blackbodies with a mass ratio of 1/10 based on fits to 70 and 160~$\mu$m data, 160 and 250~$\mu$m data, 250 and 350~$\mu$m, and 350 and 500~$\mu$m data.  Results from this analysis are shown in Figure~\ref{f_masstest_multtc}.  Understandably, the 70 and 160~$\mu$m bands produce the least reliable masses because they sample the Wien side of the dust SED and may miss most of the dust mass.  The 350 and 500~$\mu$m bands produce the best dust masses in this test.  In practice, though, the 350/500~$\mu$m ratios measured by {\it Herschel} tend to be noisy compared to the intrinsic colour variations within individual galaxies \citep[e.g. ][]{bendo12a}, and analyses based on these data would be limited by the 36~arcsec beam at 500~$\mu$m.  The 160 and 250~$\mu$m data could also be used to estimate dust masses but are more likely to underestimate the dust masses by $\gtsim$2$\times$ over a broader range of temperatures.  The dust masses based on the {\it Herschel} 250 and 350~$\mu$m data provide more reliable measurements than the masses from the 160 and 250~$\mu$m data but still have a reasonably good angular resolution for use in analyzing nearby galaxies.

Overall, this analysis shows that, for a wide range of different dust SEDs, the 250 and 350~$\mu$m data by themselves should yield reasonable dust masses within 20\% of the actual values (equivalent to a change of $<$0.1 in logarithm space).  The primary scenarios where the 250 and 350~$\mu$m data will not produce reliable dust masses are either where a substantial mass of cold dust is present but produces little emission at 350~$\mu$m or where the dust emissivity index $\beta$ is variable.  However, as we have explained, these are situations where any method for calculating dust mass will be problematic.  Dust masses based on the 250 and 350~$\mu$m data should be reasonably reliable estimates of the actual dust masses for our analysis and potentially for other analyses that do not call for complete radiative transfer modelling.

\label{lastpage}

\end{document}